\documentclass[a4paper,12pt]{article}

\PassOptionsToPackage{table}{xcolor}
\usepackage{jheppub} 
  \usepackage{rotating}                   

\usepackage[T1]{fontenc} 

\usepackage[all]{xy}
\usepackage{cancel}
\usepackage[percent]{overpic}
\usepackage{slashed}
\usepackage{wrapfig}
\usepackage{tabu}
\usepackage{diagbox}
\usepackage{mathrsfs,amsmath,amssymb,amsthm,amsfonts,tikz,graphicx,hyperref, color}
\usepackage{dsfont,epiolmec, latexsym, comment}
\usepackage{slashed,ccaption}
\usepackage{mathrsfs, calligra}
\usepackage{leftidx}
\usepackage{import}
\usepackage{multirow}
\usepackage{amsfonts}
\usepackage{pifont}
\usepackage{tabularx}
\usepackage[utf8]{inputenc}
\usetikzlibrary{intersections,calc}
\usepackage{tikz-3dplot}
\usepackage{ifthen}
\usetikzlibrary{arrows,arrows.meta}
\usepackage{amsmath}
\usepackage{xcolor}

\usepackage{caption} 
\DeclareMathOperator{\extdm}{d}
\newcommand{\extd}{\extdm \!}

\usepackage{array}
\usepackage{amsmath}

\usepackage[percent]{overpic}
\usepackage{wrapfig}
\usepackage{bbm}
\usepackage{tabu}
\usepackage{slashed}
\usepackage{fancyhdr} 
\usepackage{amsmath}
\usepackage{amsfonts}
\usepackage{amssymb}
\usepackage{diagbox}

\usepackage{multirow}
\usepackage{pifont}

\hypersetup{ linktoc=all,
    colorlinks, linkcolor={blue}, 
    citecolor={red}, urlcolor={darkpink}
}

\graphicspath{{Images/}}

\definecolor{light gray}{RGB}{220,220,220}
\definecolor{dark purple}{RGB}{108,0,217}
\definecolor{pink}{RGB}{190,20,100}
\definecolor{orang}{RGB}{193,63,0}
\definecolor{green}{RGB}{11,98,17}
\definecolor{darkpink}{RGB}{153,0,76}
\definecolor{bluegreen}{RGB}{0,102,102}
\definecolor{greenlagan}{RGB}{0,102,0}
\definecolor{redgreen}{RGB}{102,102,0}
\definecolor{Redgreen}{RGB}{153,76,0}
\definecolor{vividviolet}{rgb}{0.62, 0.0, 1.0}
\definecolor{amaranth}{rgb}{0.9, 0.17, 0.31}
\definecolor{palatinateblue}{rgb}{0.15, 0.23, 0.89}
\definecolor{brightpink}{rgb}{1.0, 0.0, 0.5}
\definecolor{cornflowerblue}{rgb}{0.39, 0.58, 0.93}
\definecolor{deepcarminepink}{rgb}{0.94, 0.19, 0.22}
\definecolor{radicalred}{rgb}{1.0, 0.21, 0.37}
\usepackage{graphicx}
\usepackage{tikz}
\usepackage{float}
\usetikzlibrary{arrows,chains,shapes,matrix,positioning,scopes}
\usetikzlibrary{decorations.markings}
\tikzstyle arrowstyle=[scale=1]
\tikzstyle directed=[postaction={decorate,decoration={markings,
    mark=at position .65 with {\arrow[arrowstyle]{stealth}}}}]
\tikzstyle reverse directed=[postaction={decorate,decoration={markings,
    mark=at position .65 with {\arrowreversed[arrowstyle]{stealth};}}}]
\usepackage{import}

\usepackage{mathrsfs,amsmath,amssymb,slashed}
\usepackage{multirow,multicol}
\usepackage{enumitem}
\usepackage[percent]{overpic}
\usepackage{slashed}

\usepackage{wrapfig}
\usepackage{tabu}
\usepackage{diagbox}
\usepackage{comment}
\usepackage{tikz}
\usepackage{mathrsfs,amsmath,amssymb,amsthm,amsfonts,graphicx,hyperref,color}
\usepackage{leftidx}
\usepackage{import}
\usetikzlibrary{decorations.pathmorphing}
\DeclareFontFamily{OT1}{rsfs}{}

\DeclareFontShape{OT1}{rsfs}{m}{n}{ <-7> rsfs5 <7-10> rsfs7 <10->rsfs10}{} 

\DeclareMathAlphabet{\mycal}{OT1}{rsfs}{m}{n}

\newcommand{\nn}{\nonumber}

\newcommand{\be}{\begin{equation}}
\newcommand{\ee}{\end{equation}}

\usepackage{orcidlink}

    \title{{\LARGE{Scaling Symmetry and Carrollian Gravity}}}
\author[]{Hamid Afshar \orcidlink{0000-0001-7786-709X}, Mehdi Ahmadi-Jahmani \orcidlink{0009-0008-0637-8764}}

\affiliation{\it Department of Physics, Faculty of Science, Ferdowsi University of Mashhad, Mashhad, Iran}

 \emailAdd{hr.afshar@um.ac.ir, ham.afshar@gmail.com, mehdiahmadijahmani@gmail.com}

\abstract{
We formulate matter-coupled scaling-Carroll gravity as a gauge theory and develop a conformal construction for generating local Carroll-invariant couplings. The construction relaxes Carrollian special conformal symmetry and leads to an enlarged gravity multiplet. After fixing the scaling symmetry, the theory is governed by the trace of the extrinsic curvature, the Carroll boost symmetry, a vector field descending from dilatation, and a symmetric tensor originating from the Carroll boost connection. The vector field acquires a St\"uckelberg-like transformation under Carroll boosts proportional to the trace of the extrinsic curvature. We show that appropriate gauge choices and field redefinitions give rise to Carrollian and Aristotelian descriptions, as well as a tensor-gauge description of the symmetric tensor field coupled to Aristotelian geometry. In the latter description, the Carroll boost parameter plays the role of a vector-valued gauge parameter.

}    
\dedicated{This work is dedicated to the memory of martyred school girls of Minab.}
\begin{document}
 \maketitle
\section{Introduction}

Carrollian geometry is the  geometric structure induced on null hypersurfaces of Lorentzian spacetimes \cite{Henneaux:1979vn,Hartong:2015xda,Ciambelli:2025unn}. Any null slice carries a degenerate temporal metric and a spatial metric whose combination defines a Carroll manifold \cite{Duval:2014uoa,Duval:2014lpa,Herfray:2021qmp,Ciambelli:2025unn}. This makes Carrollian geometry an essential ingredient in several areas of gravitational physics. Most notably, boundary field theories appearing in flat space holography naturally couple to a Carrollian background, positioning Carroll geometry as the appropriate  framework for understanding holography in asymptotically flat spacetimes \cite{Duval:2014uva,Nguyen:2023vfz,Bagchi:2023cen,Donnay:2023mrd}. The interest in this setting is further strengthened by the fact that the Carroll algebra admits infinite-dimensional conformal extensions \cite{Duval:2014uoa,Afshar:2024llh,Despontin:2025dog}, which arise as the asymptotic symmetry algebras of asymptotically flat spacetimes \cite{Duval:2014uva}. A closely related setting arises at black hole event horizons, which themselves are null hypersurfaces \cite{Donnay:2019jiz,Ecker:2023uwm,Bagchi:2023cfp}. 
Beyond gravitational applications, Carrollian structures have also emerged in field theoretic and condensed matter contexts. In particular, Carrollian kinematics appear in descriptions of fractonic matter
\cite{Bidussi:2021nmp,Baig:2023yaz,Kasikci:2023tvs,Figueroa-OFarrill:2023vbj,Figueroa-OFarrill:2023qty,Ahmadi-Jahmani:2025iqc}.

Our work focuses on the systematic construction of Carrollian local invariants, providing a natural framework for gravitational dynamics in Carrollian geometry. A useful technique for constructing such invariants is the conformal method, which is based on gauging a conformal extension of a kinematical Lie algebra (Carroll algebra in this case) and introducing compensating multiplets that transform under conformal transformations. By gauge fixing some components of the compensator, one eliminates redundant conformal symmetries and obtains the desired invariant. 
A similar method has also been applied successfully in non-relativistic contexts, such as Newton-Cartan \cite{Afshar:2015aku,Abedini:2019voz} and Hořava-Lifshitz gravity \cite{Afshar:2015aku,Devecioglu:2018apj}, and more recently in the Carrollian and Aristotelian contexts \cite{Bergshoeff:2024ilz,Bergshoeff:2025qtt}. 

 Carroll gravity theories can be derived from the ultra-relativistic limit of General Relativity (GR), where the speed of light is taken to zero \cite{Dautcourt:1997hb,Hartong:2015xda,Bergshoeff:2017btm}. When this limit is applied to the Einstein-Hilbert action, it gives rise to two distinct sectors: the electric Carrollian limit and the magnetic Carrollian limit \cite{Henneaux:2021yzg,Guerrieri:2021cdz}.
These two sectors have  been recovered as the leading and the next-to-leading order of
the Einstein-Hilbert action in powers of $c$, respectively \cite{Dautcourt:1997hb,Hansen:2021fxi}. 
Carrollian theories of gravity can also appear from gauging the Carroll algebra \cite{Hartong:2015xda,Bergshoeff:2017btm,Guerrieri:2021cdz,Figueroa-OFarrill:2022mcy}, which similarly yields electric and magnetic sectors
\cite{Campoleoni:2022ebj}. 
The distinction between these two sectors is geometrically encoded in the intrinsic torsion of the Carroll structure $(\tau^\mu, h_{\mu\nu})$
\cite{Figueroa-OFarrill:2020gpr}. The intrinsic torsion of the Carroll structure, in the magnetic Carroll gravity has to be zero  and thus the space is absolute \cite{Campoleoni:2022ebj}, whereas electric Carroll gravity corresponds to a torsional phase where the space can be dynamical \cite{Hartong:2015xda,Musaeus:2023oyp}. 
Our Carrollian conformal construction is based on gauging the \emph{anisotropic scaling Carroll} symmetry and allows for both electric and magnetic gravity sectors as solutions.

Our conformal construction couples a real compensating scalar $\phi$ to scaling-Carroll gravity, preserving local Carroll and $z$-scaling symmetries while excluding Carrollian special conformal transformations (SCTs). This relaxation of symmetry constraints is crucial, as it allows for the introduction of an additional spatial vector $b_a$ originating from the dilatation gauge field. After gauge fixing $\phi=1$, which fixes the dilatation symmetry and yields a Carrollian gravity theory, this vector acquires a shift transformation under Carroll boosts proportional to the trace of the extrinsic curvature $K$. As a result, when $K\neq0$, one can construct boost-invariant combinations and access more general torsional geometries than in standard conformal Carroll frameworks.

Depending on how the boost symmetry and the residual vector field $b_a$ are treated after fixing the scale symmetry, our construction leads to different geometric descriptions. In particular, when $K\neq0$, fixing the boost symmetry via the gauge condition $b_a=0$ leads to an Aristotelian description, while keeping the boost symmetry manifest allows the construction of boost-invariant variables through a St\"uckelberg mechanism. In this case, the independent symmetric tensor field admits a tensor-gauge reinterpretation on an Aristotelian background, with the Carroll boost parameter playing the role of a vector-valued gauge parameter. This provides a unified framework for Carrollian gravity, Aristotelian geometry, and tensor-gauge structures within a single geometric setup. The relations between these descriptions are summarized in Fig.~\ref{fig1}.

It should be emphasized that both the Carrollian and Aristotelian descriptions appearing here are equipped with an additional symmetric tensor field $S_{ab}$ originating from the Carroll boost connection. The tensor-gauge description discussed in this work may be viewed as a reinterpretation of this same field after suitable field redefinitions.

\begin{figure}[H]
	\centering
	\resizebox{0.92\linewidth}{!}{%
		\begin{tikzpicture}[
			box/.style={
				draw,
				rounded corners,
				minimum width=3.2cm,
				minimum height=1.4cm,
				align=center,
				thick
			},
			arrow/.style={
				->,
				thick,
				>=latex
			}
			]
			
			\node[box] (scgt)
			{\bfseries Scaling-Carroll Gauge Theory};
			
			\node[box, below=2.8cm of scgt] (dcg)
			{\bfseries Torsional Carroll Gravity\\ ($K\neq0$)};
			
			\node[box, left=3.5cm of dcg] (ag)
			{\bfseries Aristotelian Gravity};
			
			\node[box, right=3.5cm of dcg] (tg)
			{\bfseries Vector-Charge\\ Tensor Gauge Theory};
			\draw[arrow] (scgt) --
			node[midway, right] {$\phi=1$}
			(dcg);
			
			\draw[arrow] (dcg) --
			node[midway, above] {$b_a=0$}
			(ag);
			
			\draw[arrow] (dcg) --
			node[midway, above] {$b_a\neq0$}
			(tg);
			
		\end{tikzpicture}%
	}
	\caption{\small{
			Schematic relation between different descriptions arising from scaling-Carroll gauge theory.
			Fixing the dilatation symmetry by $\phi=1$ yields Carrollian gravity.
			For the $K\neq0$ sector, imposing $b_a=0$ gauge fixes the Carroll boost symmetry and leads to an Aristotelian description.
			Alternatively, retaining the St"uckelberg field $b_a$ allows the construction of boost-invariant Aristotelian variables and a tensor-gauge reinterpretation of the independent symmetric tensor $S_{ab}$, with the Carroll boost parameter $\lambda_a$ playing the role of a vector-valued gauge parameter suggesting a geometric realization of a vector-charge tensor gauge structure.
		}}
	\label{fig1}
\end{figure}

%
%
%
%
%
%
%
%
%
%
%
%

The remainder of this paper is organized as follows. In Section~\ref{sec:Carrollian-geometry}, we review the geometric structure of Carrollian manifolds, including the role of the extrinsic curvature and torsional extensions that naturally arise in the ultra-relativistic limit. Section~\ref{sec:Scaling-Carroll-algebra} presents the scaling Carroll algebra and its gauging, establishing the framework for implementing local Carroll and scale symmetries. In Section~\ref{sec:conformal-construction}, we introduce the conformal construction with a compensating scalar field, which allows the systematic generation of Carrollian local invariants.  Then we analyze the different  geometric description --- Carrollian, Aristotelian, and tensor-gauge theory --- that arise.
Section~\ref{sec:Carrollian-gravity} develops the corresponding Carrollian gravity theory, including scaling-Carroll field theories and curvature invariants. Finally, Section~\ref{sec:Conclusion} summarizes our results and outlines possible directions for future research.
\paragraph{Notation and convention.} We work in $D = d + 1$ space-time dimensions, where $d$ refers to the number of spatial dimensions. The small Latin alphabet letters $(a, b, c, \cdots)$ refer to the spatial local Carroll frame. The Greek indices $(\mu, \nu, \rho, \cdots)$ refer to the coordinate frame and labels all spacetime coordinates $(x^\mu\equiv t, x^i)$ where $i=1,\cdots,d$. We sometimes use $A\cdot B$ to represent the contraction of the spatial indices in $A$ and $B$. 

\section{Review of Carroll geometry}\label{sec:Carrollian-geometry}
The Carrollian structure in a $(d+1)$-dimensional spacetime is described by a degenerate metric $h_{\mu\nu}$ of rank $d$ and the nowhere-vanishing unit vector field  $\tau^\mu$.
The kernel of this tensor field is generated by the nowhere-vanishing vector field $\tau^\mu$
\begin{align}
	h_{\mu\nu}\tau^{\mu}=0\,.
\end{align}
The  inverse tensors $\tau_\mu$ and $h^{\mu\nu}$ can be defined through the following orthonormality and completeness relations
\begin{align}\label{orhtocomplet}
	\tau_{\mu}\tau^{\mu}=1\,,\qquad \tau_{\mu}h^{\mu\nu}=\tau^{\mu}h_{\mu\nu}=0\,,\qquad \delta_{\mu}^{\nu}=\tau_{\mu}\tau^{\nu}+h^{\nu\rho}h_{\rho\mu}\,,
\end{align}
where $\tau_\mu$ is a nowhere-vanishing covector that defines the temporal direction. Thus, $h_{\mu\nu}$
acts as a projector onto the spatial directions orthogonal to it. 
In terms of the co-frame form fields $e_{\mu}{}^{a}$ we have $h_{\mu\nu}=\delta_{ab} e_{\mu}{}^{a} e_{\nu}{}^{b}$ and $h^{\mu\nu}=\delta^{ab}e^\mu{}_ae^\nu{}_b$ where $a=1,\cdots,d$. The relations \eqref{orhtocomplet} are
\begin{align}\label{coframe}
	\tau^\mu\tau_\mu=1\,,\qquad e_{\mu}{}^{a}e^{\mu}{}_{b}=\delta^{a}_{b}\,, \qquad \tau^{\mu}e_{\mu}{}^{a}=0=e^{\mu}{}_{a}\tau_{\mu}\,,\qquad e_{\mu}{}^{a}e^{\nu}{}_{a}=\delta_{\mu}^{\nu}-\tau^{\nu}\tau_{\mu}\,.
\end{align}

The vector $\tau^\mu$ and the metric $h_{\mu\nu}$ are by definition invariant under local homogeneous Carroll transformation. In fact, under a general coordinate transformation with
parameter $\xi^\mu$ and local spatial rotation transformations with parameters $\lambda^{ab}$ we have \cite{Hartong:2015xda}:
\begin{subequations}
\begin{align}
	{\delta} \tau^{\mu}&=\xi^\nu\partial_\nu\tau^\mu-\tau^\nu\partial_\nu\xi^\mu\,,\\
	{\delta} e_{\mu}{}^{a}&=\xi^\nu\partial_\nu e_\mu{}^a+\partial_\mu\xi^\nu e_\nu{}^a+\lambda^{a}{}_{b}e_{\mu}{}^{b}\,.
\end{align}
\end{subequations}
From the orthonormality relations of  vielbein in \eqref{orhtocomplet} we can simply identify the transformation of $\tau_\mu$ and $e^\mu{}_a$ in which we see redundancies of the form $\delta\tau_\mu=\cdots+\lambda_ae_\mu{}^a$ and $\delta e^\mu{}_a=\cdots-\lambda_a\tau^\mu$ where the signs are set such that $\tau_\mu e^\mu{}_a=0$. We identify this new parameter $\lambda^a$, as the Carroll boost transformation. We thus have
\begin{subequations}	
	\begin{align}
		{\delta} \tau_{\mu}&=\xi^\nu\partial_\nu \tau_\mu+\partial_\mu\xi^\nu \tau_\nu+\lambda_ae_\mu{}^a\,,\\
		{\delta} h^{\mu\nu}&=\xi^\sigma\partial_\sigma h^{\mu\nu}-2h^{\sigma(\mu}\partial_\sigma \xi^{\nu)}-2\lambda^a\tau^{(\mu}e^{\nu)}{}_a\,.
	\end{align}
\end{subequations}

We can define a boost invariant Carrollian expansion associated to the metric $h_{\mu\nu}$ as its Lie derivative along the vector $\tau^\mu$
\begin{align}\label{extrinsiccurv}
		K_{\mu\nu}\equiv-\frac{1}{2}\mathcal{L}_{\tau}h_{\mu\nu}=-\frac{1}{2}(\tau^{\rho}\partial_{\rho}h_{\mu\nu}+h_{\mu\rho}\partial_{\nu}\tau^{\rho}+h_{\nu\rho}\partial_{\mu}\tau^{\rho})\,.
	\end{align}
	In particular, $K_{\mu\nu}=0$ implies that the spatial metric $h_{\mu\nu}$ does not evolve along the temporal direction defined by $\tau^\mu$. In this case, the Carrollian spatial geometry is frozen in time. Conversely, a non-vanishing $K_{\mu\nu}$  signals a dynamical evolution of spatial slices and can be interpreted as a manifestation of torsional effects in Carrollian geometry \cite{Figueroa-OFarrill:2020gpr}. We may consider a Carrollian manifold as a codimension-1 null hypersurface $\Sigma$ embedded
	in a Lorentzian spacetime with $h_{\mu\nu}$ as the induced (degenerate) metric on it. A Carroll boost
	invariant measure of how the hypersurface is curved is encoded in the variation of the induced
	degenerate metric $h_{\mu\nu}$ along the normal vector $\tau^\mu$. This defines the second fundamental form,
	or extrinsic curvature, of $\Sigma$. The term `extrinsic curvature' is used for $K_{\mu\nu}$ from this perspective. 
	
	The extrinsic curvature \eqref{extrinsiccurv} can be written in terms of the co-frame fields as
	follows
	\begin{align}
		K_{\mu\nu}=-(\tau^{\rho}e_{\mu}{}^a\partial_{[\rho} e_{\nu]a}+\tau^{\rho}e_{\nu a}\partial_{[\rho} e_{\mu]}{}^{a})\,.
	\end{align}
	Having defined the anholonomy coefficients $\Omega_{\mu\nu}{}^a\equiv2\partial_{[\mu}e_{\nu]}{}^{a}$, we have\footnote{We use the vector $\tau^\mu$ and the inverse vielbein  and $e^\mu{}_a$ to turn the curved indices into flat ones. {In general for a form field $X_\mu$ we have,
			\begin{align}
				X_0=\tau^\mu X_\mu\,,\qquad X_a=e^\mu{}_a X_\mu\,, \qquad X_\mu=X_0\tau_\mu+X_a e_\mu{}^a\,.\label{projfield}
		\end{align}} 
	}
	\begin{align}\label{extrinanhol}
		K_{\mu\nu}=-e_{(\mu}{}^ae_{\nu)}{}^b\Omega_{0ab}\,.
	\end{align}
	where $\Omega_{0a}{}^c=\tau^\mu e^\nu{}_a \Omega_{\mu\nu}{}^c$ 
	denotes the temporal projection of the anholonomy coefficients, which captures the extrinsic curvature. Flat spatial indices $a,b$ are raised and lowered with $\delta_{ab}$ so we don’t need to distinguish up versus down in them. 
	A short computation further shows that $K_{\mu\nu}$ is purely spatial, that is, $\tau^{\mu}K_{\mu\nu}=0$ and thus $K_{ab}=e^\mu{}_ae^\nu{}_bK_{\mu\nu}=-\Omega_{0(ab)}$ has the same information as $K_{\mu\nu}$. We can use $h^{\mu\nu}$ to raise curved indices of purely spatial tensors like $K_{\mu\nu}$, 
	\begin{align}
		K^{\mu\nu}=h^{\mu\rho}h^{\nu\sigma}K_{\rho\sigma}
		=-e^{(\mu |a}e^{\nu) b}\Omega_{0ab}\,.
	\end{align}The trace of the extrinsic curvature is also given by 
	$K = h^{\mu\nu} K_{\mu\nu}=-\Omega_{0 a}{}^a$.
	\subsection{Carrollian scaling transformation}
	Here in this paper we discuss scaling Carroll gravity in which local scale transformation (Weyl transformation) are also present as a new gauge symmetry. In the Carolinian setup, which is non-Lorenzian, the asymmetry between time and space suggests the possibility of an anisotropic behaviour for time and space under a generic local scaling;
	\begin{align}\label{Weyl}
	\delta \tau^\mu=-z\lambda_D\tau^\mu\,,\qquad \delta h_{\mu\nu}=2\lambda_D h_{\mu\nu}\,.
	\end{align}
	Here $z$ is the anisotropic scaling parameter and $\lambda_D(\vec x,t)$ denotes the gauge parameter of Weyl transformation. 	A simple calculation gives the Weyl transformation of the extrinsic curvature; 
	\begin{align}\label{weylextr}
\delta K_{\mu\nu}=(2-z)\lambda_{D}K_{\mu\nu}-h_{\mu\nu}\partial_{0}\lambda_{D}\,,
	\end{align}  where $\partial_0=\tau^\mu\partial_\mu$.
	\subsection{Gravity-coupled single scalar field}
	As a warm-up let us try to construct a local scale invariant dynamical action using  the extrinsic curvature.   
	Since $K_{ab}$ contains the same information  $K_{\mu\nu}$ does, and transforms only under rotation as $\delta K_{ab}=2\lambda_{(a}{}^cK_{b)c}$, we have any scalar constructed from it such as $K$ and $K_{\mu\nu}^2=4K_{ab}K^{ab}$, obviously being Carroll invariant. 
	Thus with two time derivatives, we have Carroll diffeomorphism invariants constructed from the extrinsic curvature as;
	\begin{align}\label{Kinv}
		S_1=\int\extd^{d+1}x\,e\,\left(a_1\,K^2+a_2\, K_{\mu\nu}K^{\mu\nu}\right)\,,
	\end{align}
	We notice that the invariant Lagrangian $\mathcal L= e\,\tau^\mu\partial_\mu K$ is equivalent to $\mathcal L= e\,K^2$ up to a total derivative --- see appendix \ref{appA}.
	The corresponding action constructed from these invariants should be dimensionless and invariant. 
	
	We can couple a single real scalar field to the Carroll geometry with two time derivatives as follows
	\begin{align}
		S=\int\extd^{d+1}x\,e\,\left(c_1(\tau^\mu\partial_\mu\phi)^2+c_2\,K\phi\tau^\mu\partial_\mu\phi+a_1\,K^2\phi^2+a_2\, K_{\mu\nu}K^{\mu\nu}\phi^2\right)\,.
	\end{align}
	Any other Lagrangian combination like $\mathcal L=e\phi\tau^\mu\tau^\nu\nabla_\mu\partial_\nu\phi$ is equivalent to these terms up to total derivatives. 
	A curious question is whether we can make this Carroll invariant action also invariant under local scale symmetry with the transformation \eqref{Weyl} and
	\begin{align}\label{scalingtrans}
\delta\phi=w\lambda_D\phi\,.
	\end{align}
 Recalling the fact that $\delta e=(d+z)\lambda_De$ we can calculate the transformation of the action $S$ under dilatation. The homogeneous term of the transformed action will be canceled provided that the weight of the scalar field is $w=\frac{z-d}{2}$ and using \eqref{weylextr} the inhomogeneous term will be
	\begin{align}
		\delta S=\int\extd^{d+1}x\,e\left[(-2a_1d-2a_2+wc_2)K\phi^2+(2wc_1-c_2d)\phi\partial_0\phi\right]\partial_0\lambda_D\,.
	\end{align}
	We can cancel this term by adjusting the coefficients $a_2$ and $c_2$. It is interesting that unlike the relativistic case we can make the pure gravity theory $S_1$ scale invariant by setting $a_{2}=-da_1$ and $c_1=c_2=0$. If we include the scalar field Kinetic term, and after rescaling the scalar field to canonically normalize the Kinetic term by setting $c_1=\frac12$, we have the action $S$ invariant if
	\begin{align}
		c_2=\frac{z-d}{2d}\,,\qquad a_{2}=-da_1+\frac{(z-d)^{2}}{8d}\,.
	\end{align}
	The fact that the combination $a_2+da_1$ is fixed shows that we have some arbitrariness in conformally coupling the scalar field to gravity invariants in $S_1$. 
	We can rephrase these options as follows
	\begin{align}
		S        =\frac{1}{2}\int\extd^{d+1}x\,e\left[ \left( D_{0}\phi \right)^{2}+\alpha\left(K^2-d\,K_{\mu\nu}^2\right)\phi^2\right],
	\end{align}
	where $D_0=\partial_0+w\,d\, K$ and $\alpha$ is an arbitrary parameter. 
	As discussed in \cite{Bergshoeff:2024ilz}, this theory presents a conformal construction of Carrollian gravity invariants at $z=1$.
	Here, we extend this framework to any dynamical exponent $z$, starting by reviewing scaling-Carroll gravity from a gauge theory perspective to set up our conformal construction.

\section{Scaling Carroll algebra and gauging}\label{sec:Scaling-Carroll-algebra}
\label{section3}

In this section we focus on the concept of the `gravity as a gauge theory' and  aim to find the gravity theory which is invariant under local scaling Carroll spacetime transformations. 

In the  Carroll algebra, the Hamiltonian as the generator of time translation $H$ appears as the central term in the commutator of Carroll boosts $G_a$ and space translations $P_a$, forming the $d$ dimensional Heisenberg algebra;
\begin{align}\label{CarrollAlg1}
	[{P_a},{G_b}] &=\delta_{ab}{H}\,.
\end{align}
The rest of non-zero commutators constitute spatial rotation generators $J_{ab}$ acting as the subalgebra $so(d)$;
\begin{align}\label{CarrollAlg2}
	[J_{ab},{P_c}]&=2\delta_{c[a}P_{b]} \,,\quad\quad [{J_{ab}},{G_c}]=2\delta_{c[a}G_{b]}\,,\quad\quad [{J_{ab}},{J_{cd}}]=4\delta_{[a[d}J_{c]b]}\,.
\end{align}
The scaling extension of the Carroll algebra is the one-parameter family $\mathfrak{scalcarr}_z(d+1)$ in $d+1$ spacetime dimensions \cite{Afshar:2024llh}, labeled by the parameter $z$, which plays the role of a dynamical exponent analogous to that in the Lifshitz algebra. These algebras are generated by time translations $H$, spatial translations $P_a$, Carroll boosts $G_a$, rotations $J_{ab}$, and dilatations $D$, with the following non-vanishing commutators:\footnote{ Our convention compares to \cite{Bergshoeff:2024ilz} as;
	\begin{align}
		P_{A}\to P_a\,,\quad J_{0A}\to G_a\,,\quad J_{AB}\to -J_{ab}\,,\quad D\to -D\,,\quad H\to -H\,, \quad z\to1\,.\nn
\end{align}}
\begin{align}\label{CarrollAlg3}
	\begin{split}
		[{D},{H}]&=-z{H},\quad[{D},{P}_{a}]=-{P}_{a}\,,\quad [{D},{G}_{a}]=(1-
		z){G}_{a}\,.
	\end{split}
\end{align}
This algebra admits an extension with the generator of temporal SCT; $C$, by including two non-vanishing commutators $[D,C]=(2-z)C$ and $[C,P_a]=-2G_a$ for generic values of $z$, denoted as $\mathfrak{confcarr}_z(d+1)$ in \cite{Afshar:2024llh}. However, including both temporal and spatial SCT generators requires $z=1$, in which case the algebra is usually referred to as the Carrollian conformal algebra. The focus of the present work is $\mathfrak{scalcarr}_z(d+1)$ defined by the commutation realtions \eqref{CarrollAlg1}-\eqref{CarrollAlg3}.

\subsection{Gauging the scaling Carroll algebra}
Starting with the scaling Carroll algebra  \eqref{CarrollAlg1}-\eqref{CarrollAlg3}, in this section we develop the gauging procedure on which we base our conformal method. We do this by associating a connection gauge field $A_{\mu}$ with all generators of the scaling Carroll algebra and a gauge transformation $\Lambda$ as
\begin{align}\label{connection}
	A_{\mu}&={H}\,\tau_{\mu}+{P}_{a}\,e_{\mu}{}^{a}+{G}_{a}\,\omega_{\mu}{}^{a}+\frac{1}{2}{J}_{ab}\,\omega_{\mu}{}^{ab}+{D}\,b_{\mu}\,,\\\label{gaugetransf}
	\Lambda&=\frac{1}{2}{J}_{ab}\,\lambda^{ab}+{G}_{a}\,\lambda^a+{D}\,\lambda_{D}\,.
\end{align}
The parameters $\lambda^{ab}$, $\lambda^{a}$ and $\lambda_{D}$ are local rotation, local Carroll boost, and local dilatation transformations. The transformation of the various gauge fields in \eqref{connection} under  gauge symmetries of \eqref{gaugetransf} can be compactly written as $\delta_{\text{\tiny{gt}}} A_\mu=\partial_{\mu}\Lambda+[A_{\mu},\Lambda]$.
Using the scaling Carroll algebra we can easily derive the transformation rules as follows
\begin{subequations}\label{homtransf}
	\begin{align}
		{\delta}_{\text{\tiny{gt}}}\tau_{\mu} &=\lambda_{a}e_{\mu}{}^{a}+z\lambda_{D}\tau_{\mu},\\
		{\delta}_{\text{\tiny{gt}}}e_{\mu}{}^{a} &=\lambda^a{}_be_{\mu}{}^{b}+\lambda_{D}e_{\mu}{}^{a},\\
		{\delta}_{\text{\tiny{gt}}}\omega_{\mu}{}^{a} &=\partial_{\mu}\lambda^{a}+\lambda^{a}{}_{b}\omega_{\mu}{}^{b}-\lambda_{b}\omega_\mu{}^{a b}+(1-z)\lambda^{a}b_{\mu}+(z-1)\lambda_{D}\omega_{\mu}{}^{a}, \\
		{\delta}_{\text{\tiny{gt}}}\omega_{\mu}{}^{ab} &=\partial_{\mu}\lambda^{ab}+\lambda^{a}{}_{c}\omega_{\mu}{}^{cb}-\lambda^{b}{}_{c}\omega_{\mu}{}^{ca}\,,\\
		{\delta}_{\text{\tiny{gt}}}b_{\mu}&=\partial_{\mu}\lambda_{D}\,.
	\end{align}
\end{subequations}
The gauge covariant curvature $F_{\mu\nu}$ is defined as  
$F_{\mu\nu}=\partial_{\mu}A_{\nu}-\partial_{\nu}A_{\mu}+[A_{\mu},A_{\nu}]$ which obeys Bianchi identities, \begin{align}\label{Bianchiiden}
	D_{[\lambda}\, F_{\mu\nu]}=\partial_{[\lambda}\, F_{\mu\nu]}+[A_{[\lambda},\, F_{\mu\nu]}]=0\,,
\end{align}
where  a complete anti-symmetrization among free indices are understood.  The curvature of the gauge field of the Carroll $z$-rescaling algebra \eqref{connection} can be expanded as 
\begin{align}\label{fieldstrength}
	F_{\mu\nu}={H}\,R_{\mu\nu}(H)+{P}_{a}\,R_{\mu\nu}{}^{a}(P)+\frac{1}{2}{J}_{ab}\,R_{\mu\nu}{}^{ab}(J)+{G}_{a}\,R_{\mu\nu}{}^{a}(G)+{D}\,R_{\mu\nu}(D)\,.
\end{align}
These components are given as,
\begin{subequations}
	\begin{align}\label{curvature 2-form}
		R_{\mu\nu}(H)&=2\partial_{[\mu}\tau_{\nu]}-2\omega_{[\mu}{}^{a}e_{\nu]a}+2z\tau_{[\mu}b_{\nu]}\,,\\
		R_{\mu\nu}{}^{a}(P)&=2\partial_{[\mu}e_{\nu]}{}^{a}-2\omega_{[\mu}{}^{ab}e_{\nu]b}+2 e_{[\mu}{}^{a}b_{\nu]}\,,\\
		R_{\mu\nu}{}^{a}(G)&=2\partial_{[\mu}\omega_{\nu]}{}^{a}-2\omega_{[\mu}{}^{ab}\omega_{\nu]b}+2(z-1)\omega_{[\mu}{}^{a}b_{\nu]}\,,\\
		R_{\mu\nu}{}^{ab}(J)&=2\partial_{[\mu}\omega_{\nu]}{}^{{ ab}}-2\omega_{[\mu}{}^{c[a}\omega_{\nu]}{}^{b]}{}_c\,,\\
		R_{\mu\nu}(D)&=2\partial_{[\mu}b_{\nu]}\,.
	\end{align}
\end{subequations}
All gauge fields we have defined up to here are independent with their independent gauge transformation \eqref{homtransf}. In order to interpret this gauge theory as a gravity theory we should have some gauge fields depending on geometric independent variables such as vielbein, and we should also identify the local infinitesimal diffeomorphism with the action of local time and spital translation.
By imposing suitable curvature constraints and solving them, we will be able to obtain some of the dependent gauge field components and also interpret diffeomorphism as a gauge transformation along $H$ and $P_a$.
One could impose the following constraints \cite{Bergshoeff:2024ilz};
\begin{align}\label{constraints}
	R_{\mu\nu}(H)=0\,,\qquad R_{ab}{}^c(P)=0\,,\qquad R_{0a}{}^a(P)=0\,,\qquad R_{0[ab]}(P)=0\,.
\end{align}
Upon imposing the constraints \eqref{constraints}, additional constraints are found using the Bianchi identity \eqref{Bianchiiden}. For example, the identity along $H$, 
\begin{align}
	\begin{split}
		\partial_{[\lambda}R_{\mu\nu]}(H)- \omega_{[\lambda}{}^a R_{\mu\nu]}{}^{a}(P)+ e_{[\lambda}{}^a R_{\mu\nu]}{}^a(G)+z\,\tau_{[\lambda}R_{\mu\nu]}(D)-z\,b_{[\lambda} R_{\mu\nu]}(H)=0\,,\\
	\end{split}
\end{align}
leads to new constraints  
\begin{align}
	R_{[abc]}(G)=0\,,\qquad2\omega_{[a|c}R_{0|b]}{}^c(P)+2R_{0[ab]}(G)+z\,R_{ab}(D)=0\,.
\end{align}
where some components of $\omega_{ab}$ is now dependent as we see below.
Now we are ready to solve all constraints \eqref{constraints} one by one:
\begin{itemize}[leftmargin=*]
	
	\item{$R_{ab}(H)=0$ ---} This set of constraints determines the antisymmetric part of the spatial projection of the boost spin-connection gauge field, $\omega_{ab}=\omega_{[ab]}+\omega_{(ab)}$ while leaving the symmetric part undetermined;
	\begin{align}
		\omega_{[ab]}(\tau,e)=\frac12\tau_{ab}\,,\qquad\omega_{(ab)}=S_{ab}\,,
	\end{align}
	where  $\tau_{\mu\nu}\equiv\partial_{\mu}\tau_{\nu}-\partial_{\nu}\tau_{\mu}$ and thus $\tau_{ab}=e^\mu{}_a e^\nu{}_b \tau_{\mu\nu}$. $S_{ab}$ is an independent symmetric tensor. 
	
	\item{$R_{0a}(H)=0$} --- This set of constraints solves the temporal projection of the boost spin-connection gauge field as;
	\begin{align}
		\omega_{0a}(\tau,e,b_a)=\tau_{0a}+z\,b_{a}\,,
	\end{align}
	where  $\tau_{0a}=\tau^\mu e^\nu{}_a \tau_{\mu\nu}$. As we will see below the spatial component of the dilatation gauge field $b_a$ also remains as an independent field in our final gravity theory and will not be solved in our setup.
	In fact, this marks an important departure of our setup from that used in \cite{Bergshoeff:2024ilz}, where the presence of an additional SCT symmetry forced the field $b_a$  to be absent from their final result.
	Using the decomposition \eqref{projfield} and \eqref{coframe}  we can express the final form of the boost gauge field in terms of independent gauge fields as,
	\begin{align}\label{boostspin}
		\omega_{\mu}{}^ a(\tau,e,b)    =\tau_{\mu}\,e^{\nu  a}\tau^{\rho}\partial_{[\rho}\tau_{\nu]}+e^{\nu  a}\partial_{[\mu}\tau_{\nu]}+S^{ab}e_{\mu  b}+z \,b^{a}\tau_{\mu}\,.
	\end{align}
	\item{$R_{ab}{}^c(P)=0$ ---} These set of constraints can be used to solve for the spatial projection  
	of the rotation spin-connection $\omega_\mu{}^{ab}$.    For algebraically  solving it we can use the combination $
	R_{abc}(P)+R_{cab}(P)-R_{bca}(P)=0\,,
	$ which results in
	\begin{align}\label{spatialspinconnection}
		\omega_{acb}(e,b_a)
		=\frac12\left(\Omega_{abc}+\Omega_{cab}+\Omega_{cba}\right)+b_{b}\delta_{ac}-\delta_{ab}b_c\,,
	\end{align}
	where 
	$\Omega_{\mu\nu}{}^a\equiv2\partial_{[\mu}e_{\nu]}{}^{a}$ and thus
	$\Omega_{ab}{}^c=e^\mu{}_a e^\nu{}_b \Omega_{\mu\nu}{}^c$   are the spatial anholonomy coefficients.\footnote{A non-holonomic frame is one with non-vanishing $\Omega$s.}
In order to solve for the temporal spin-connection $\omega_0{}^{ab}$ we need an extra constraint.
\item{$R_{0[a}{}_{b]}(P)=0$ --- } Using this constraint 
we can solve for $\omega_0{}^{ab}$,
\begin{align}\label{temporalspinconnection}
	{\omega}_0{}^{ab}(\tau,e)=-2\tau^{\mu}e^{\nu  [a}\partial_{[\mu}e_{\nu]}{}^{b]}=-\Omega_0{}^{[ab]}\,.
\end{align}
The full-rank spin-connection tensor can now be obtained from \eqref{spatialspinconnection} and \eqref{temporalspinconnection} as     follows,     where the field $b_a$ again remains as an independent arbitrary  field
\begin{align}\label{rotationspin}
	\omega_{\mu}{}^{ab}(\tau,e,b)
	=-2e^\nu{}^{  [a}\partial_{[\mu}e_{\nu]}{}^{ b]}+e_{\mu  c}e^{\rho a}e^{\sigma  b}\partial_{[\sigma}e_{\rho]}{}^{ c}+2e_{\mu}{}^{ [a}b^{b]  }\,.
\end{align}
\item{$R_{0a}{}^a(P)=0$ --- } Finally the temporal projection of the $b_\mu$ gauge field is solved here
\begin{align}\label{b0gaugefield}
	b_{0}(\tau,e)=\frac{2}{d}\tau^{\mu}e^{\nu}{}_{ a}\partial_{[\mu}e_{\nu]}{}^{a}=-\frac{1}{d}K\,.
\end{align}
where in the last equality, we used the definition \eqref{extrinanhol} to express $b_0$ as an invariant geometric quantity. 
\end{itemize}
The only non-constraint component of the curvature 2-form $R_{\mu\nu}{}^{a}(P)$ is the symmetric-traceless part of $R_{0a b}(P)$  which we denote as $\cancel{R}_{0}{}^{(ab)}(P)\equiv R_{0}{}^{(ab)}(P)-\frac{1}{d}\delta^{ab}\,R_{0c}{}^{c}(P)$ and captures the information of the extrinsic curvature; 
\begin{align}\label{torsion}
\cancel{R}_{0}{}^{(ab)}(P)=\Omega_0{}^{(ab)}-b_0(\tau,e)\delta^{ab}
=-K^{ab}+\frac1d\delta^{ab}K\,.
\end{align}
This geometric quantity transforms covarianlty under dilatation as $\delta_D\cancel{R}_{0}{}^{(ab)}(P)=-z\lambda_D\cancel{R}_{0}{}^{(ab)}(P)$.
\subsection{Transformation rules}\label{sec.trns}We can divide the list of dependent and independent components of the scaling Carroll gauge fields as
\begin{center}
\begin{tabular}{  c |c }
	independent components & dependent components  \\ 
	\hline
	$e_\mu{}^a$ & $\omega_\mu{}^{ab}$  \\  
	\hline
	$\tau_\mu$ & $\omega_{0a}$    \\ 
	\hline
	$b_a$ & $b_0$     \\
	\hline
	$\omega_{(ab)}$ & $\omega_{[ab]}$     
\end{tabular}
\end{center}
The transformation of the independent fields $\tau_\mu$, $e_\mu{}^a$, $b_a$ and $\omega_{(ab)}$ naturally follow \eqref{homtransf}. The boost transformation of the independent gauge fields $\tau_\mu$ and $e_\mu{}^a$ and their inverses are given in \eqref{boosttrans}. The independent components of the dilatation and spin connection gauge fields also transform under Carroll boost as follows
\begin{align}\label{boostindpnt}
\delta_{{\text{\tiny G}}} b_a=-\lambda_{a}b_{0}\,, \qquad
\delta_{{\text{\tiny G}}} \omega_{(ab)}=D_{(a}\lambda_{b)}-\lambda_{(a}\omega_{0|b)}\,,
\end{align}
where $D_a\lambda_b=\partial_a\lambda_b-\omega_{abc}\lambda_c-(z-1)b_a\lambda_b$. Especially the spatial component $b_a=e^\mu{}_ab_\mu
$ transforms as a shift under the Carroll-boost transformation and thus should appear in the scaling Carroll gravity multiplet, this is unlike the case of conformal Carroll gravity where the presence of the special conformal Carroll symmetry, enforces it to drop out any invariant. 

Among the dependent fields, the temporal component of the dilatation gauge field  $b_0 = \tau^\mu b_\mu$ is Carroll-boost invariant and plays the role of the trace of the extrinsic curvature. 
It is essential to obtain the transformation rules of dependent fields and compare with their transformation as independent gauge fields mentioned in \eqref{homtransf}. 
We can show that these two transformations are not necessarily the same, and the transformation as a dependent gauge field can be deviated by some amount which we show by $\Delta$ --- the details of the derivation is presented in appendix \ref{trnsrules};
\begin{subequations}
\begin{align}
	\Delta \omega_{[ab]}&=\delta\omega_{[ab]}(\tau,e)-\delta_{\text{\tiny{gt}}}\omega_{[ab]}=0\,,\\
	\Delta \omega_{0a}&
	=\delta \omega_{0a}(\tau, e, b_a)-\delta_{\text{\tiny{gt}}}\omega_{0a}=\frac{1}{d}\lambda_a K-\lambda_b K_{ab}= \lambda_b\cancel{R}_{0}{}_{(ab)}(P)\,,\\
	\Delta \omega_{0ab}&=\delta \omega_{0ab}(\tau,e)-\delta_{\text{\tiny{gt}}} \omega_{0ab}=0\,,\\
	\Delta \omega_{abc}&=\delta \omega_{abc}(e,b_a)-\delta_{\text{\tiny{gt}}} \omega_{abc}=-\lambda_b\cancel{R}_{0}{}_{(ac)}(P)+\lambda_c\cancel{R}_{0}{}_{(ab)}(P)\,.
\end{align}
\end{subequations}

The same happens for the field strengths that appear in \eqref{fieldstrength}; their transformation as field strength of gauge fields follows immediately $\delta_{{\text{\tiny gt}}} F_{\mu\nu}=[F_{\mu\nu},\Lambda]$. However, this might not coincide with their transformation as filed strength of dependent fields. The non-zero field strengths transform as --- see appendix \ref{trnsrules} 
\begin{align}
\Delta \cancel{R}_{0}{}^{(ab)}(P)&=0\,,\qquad
\Delta {R}_{0a}(D)=0\,,\qquad
\Delta {R}_{ab}(D)=0\,,\nn\\
\Delta R_{0a}{}^{b}(G)&=R_{0a}(H)\lambda_{c}\cancel{R}_{0}{}^{(bc)}(P)-D_{a}(\lambda_{c}\cancel{R}_{0}{}^{(bc)}(P))+b_a\lambda_{c}\cancel{R}_{0}{}^{(bc)}(P)\,,\nn\\
\Delta R_{ca}{}^{cb}(J)&=R_{cae}(P)\lambda^{e}\cancel{R}_{0}{}^{(cb)}(P)+2D_c\left(\lambda_a\cancel{R}_{0}{}^{(cb)}(P)\right)-2b_c\lambda_a\cancel{R}_{0}{}^{(cb)}(P)\,,
\end{align}
where $D_{a}(\lambda_{b}\cancel{R}_{0}{}^{(ab)}(P))=\partial_{a}(\lambda_{b}\cancel{R}_{0}{}^{(ab)}(P))-\omega_{a}{}^{a}{}_{c}\lambda_{b}\cancel{R}_{0}{}^{(bc)}(P) +b_a(\lambda_{b}\cancel{R}_{0}{}^{(ab)}(P))$. This analysis
shows that, we can consider the following boost-invariant scalar combination from the nonzero components of the field strengths $R_{\mu\nu}(D)$ and $R_{\mu\nu}{}^a(P)$,
\begin{align}\label{curvatureinv1}
L_1=R_{0a}(D)R_{0a}(D)\,,\qquad L_2=\cancel{R}_{0}{}_{(ab)}(P)\cancel{R}_{0}{}^{(ab)}(P)\,.
\end{align}
The standard boost gauge transformation follow from \eqref{gtindepend}-\eqref{gtindepend1}. In particular we have,
\begin{align}\label{curvatureinv2}
\delta_{{\text{\tiny gt}}} R_{0ab}(G)&=-\lambda_cR_{0abc}(J)+(1-z)\lambda_bR_{0a}(D)\,,\\
\delta_{{\text{\tiny gt}}} R_{cacb}(J)&=-\lambda_cR_{0abc}(J)-\lambda_aR_{c0cb}(J)\,.
\end{align}
Upon imposing the constriants $R_{0a}(H)=0=R_{abc}(P)$, we also have an extra invariant scalar
\begin{align}
L_3=R(G,J)\equiv2R_{0a}{}^{a}(G)+R_{ab}{}^{ab}(J)+2(z-1)R_{0a}(D)M_a\,,
\end{align}
where $M_a\equiv \frac dKb_a$ -- see section \ref{sec:conformal-construction}.
We will return to this point later when discussing invariant curvature terms in section \ref{curvterms}.
The dilatation transformations for these invariants are expressed as follows:
\begin{align}
\begin{split}
	\delta_D L_1=-2(z+1)\lambda_{D}L_1\,,\qquad\delta_D L_2=-2z\lambda_{D}L_2\,,\qquad\delta_D L_3=-2\lambda_{D}L_3\,.
\end{split}
\end{align}

\section{Carrollian conformal construction}\label{sec:conformal-construction}

We augment the scaling-Carroll gravity multiplet introduced in the previous section by adding a compensating scalar field that transforms under local dilatations according to \eqref{scalingtrans}. Gauge-fixing this scalar removes the dilatation symmetry, reducing the theory to one with only local Carroll symmetries. However, as we will see, this procedure does not completely eliminate the effects of the larger symmetry: it leaves behind additional independent fields in the gravity multiplet, which play a central role in the resulting geometric and dynamical structure. The compensating scalar $\phi$, together with the scaling-Carroll vielbein gauge fields $\tau_\mu$ and $e_\mu{}^a$, transform under dilatations (with parameter $\lambda_D$) according to \eqref{Weyl}.
Upon gauge fixing the dilatation symmetry by setting the scalar field to one, $\phi \to 1$, the independent gauge fields of the scale-invariant Carroll gravity, $(\tau_\mu, e_\mu{}^a)$, reduce to the geometric variables of Carroll gravity, for which we will use the same notation for simplicity. In addition, this reduction leaves two remaining independent fields
\begin{align}\label{indepfields}
	b_a \,, \qquad
	S_{ab}\,.
\end{align}
In this context, $b_a$ is interpreted as a spatial vector field and $S_{ab}$ as a symmetric spatial tensor in the Carroll gravity. Their transformations under local Carroll boosts are given by \eqref{boostindpnt}.\begin{align}
	\delta_{\text{\tiny G}} b_a&=\frac1dK\lambda_a \,,\label{trnasV}\\
	\delta_{\text{\tiny G}}  S_{ab}&= \mathcal D_{(a}\lambda_{b)}-2(z-1)b_{(a}\lambda_{b)}-\delta_{ab}b\cdot\lambda-\lambda_{(a}\widehat\omega_{0|b)}\,.\label{indeptransf}
\end{align}
Their transformation under rotation is standard and under coordinate transformation they transform as a scalar. The  Carrollian covariant derivative $\mathcal{D}$ is defined according to the transformation of each field, i.e. 
\begin{align}\label{covderivindep}
	\mathcal D_\mu\lambda_{a}&=\partial_{\mu}\lambda_{a}-\widehat\omega_\mu{}_a{}^{b}\lambda_{b}\,,\\
	\mathcal{D}_{\mu} b_{ a}&=\partial_{\mu}b_{a}-\widehat\omega_\mu{}_a{}^{b} b_{b}+\widehat\omega_\mu{}_ab_{0}\,.
\end{align}
The Carrollian spin connections $\widehat\omega$ are naturally defined in terms of the scaling-Carroll spin connection $\omega$, addressed in section \ref{section3}, once the dilatation gauge field $b_\mu$ is zero;
\begin{align}\label{Carrollboosthat}
	\omega_{\mu}{}^a= \widehat\omega_{\mu}{}^a+z\,b^a\tau_\mu\,,\qquad\qquad    \omega_{\mu}{}^{ab}=\widehat\omega_{\mu}{}^{ab}+2e_\mu{}^{[a}b^{b]}\,.
\end{align}
According to \eqref{boostspin} and \eqref{rotationspin}, they are expressed as,
\begin{align}\label{Carrspincon}
	\widehat\omega_{\mu a}=\frac12\left(\tau_{\mu}\tau_{0}{}_a+\tau_{\mu}{}_a\right)+S^{ab}e_{\mu  b}\,,\qquad \widehat{\omega}_{\mu}{}_{ab}=-\Omega_{\mu}{}^{[a b]}+\frac12e_{\mu  c}\Omega^{ba c}\,.
\end{align}
Thus $\widehat\omega_{0 a}=\tau_{0a}$ while $\widehat\omega_{[ab]}=\frac12\tau_{ab}$ and $\widehat\omega_{(ab)}=S_{ab}$. 
When \(K \neq 0\), it is useful to introduce the vector field $M_a$, defined through its relation to $b_a$ as
\begin{equation}
	b_a = \frac{1}{d}\, K\, M_a\,,
\end{equation}
such that it transforms under Carrollian boosts as\footnote{ 
	The field $M_a$ can be the spatial projection of the contravariant vector $M^\mu$ introduced in \cite{Hartong:2015xda} as the compensator for Carroll boosts and  $\theta^\mu$ introduced in \cite{Armas:2023dcz}  in the construction of Carrollian fluids.}
\begin{equation}\label{Maboost}
	\delta_{\text{\tiny G}} M_a = \lambda_a\,.
\end{equation}
Its Carrollian covariant derivative is therefore given by
\begin{align}\label{covderM}
	\mathcal{D}_\mu M_a = \partial_\mu M_a - \widehat{\omega}_{\mu ab} M_b - \widehat\omega_{\mu a}\,,
\end{align}
where $\widehat\omega$ are given in \eqref{Carrspincon}.

At this stage, the Carroll gravity multiplet consists of the independent fields
\((\tau_\mu, e_\mu{}^a, b_a, S_{ab})\).
According to \eqref{trnasV}, the physical interpretation of the theory is largely controlled by the vector field \(b_a\), the trace of the extrinsic curvature \(K\), and the treatment of the Carroll boost parameter \(\lambda_a\).
The resulting regimes can be organized as follows:

\begin{enumerate}
	
	\item \textbf{Carrollian gravity.}
	The boost symmetry remains unfixed, i.e.\ \(\lambda_a\) is arbitrary.
	This defines a genuinely Carrollian geometry whose properties depend on the trace of the extrinsic curvature \(K\).
	
	\begin{itemize}
		
		\item \emph{Sheared Carroll gravity} (\(K=0\)):  
		The vanishing trace of the extrinsic curvature implies that $\tau^\mu$ is volume-preserving,
		\begin{align}
			h^{\mu\nu}\mathcal{L}_\tau h_{\mu\nu}=0,
		\end{align}
		and therefore the intrinsic torsion of the Carrollian structure is traceless \cite{Figueroa-OFarrill:2020gpr}. In this case,
		\begin{align}
			K_{ab}=\Theta_{ab}+\frac{1}{d}\delta_{ab}K\,, \qquad K=0,
		\end{align}
		where \(\Theta_{ab}=-\cancel{R}_{0(ab)}(P)\) is the traceless Carrollian shear tensor \cite{Donnay:2019jiz}. This defines a novel Carrollian sector which admits both magnetic (\(K_{ab}=0\)) and electric (\(K_{ab}\neq0\)) realizations.
				
		\item \emph{Torsional Carroll gravity} (\(K\neq0\)):  
		The trace of the extrinsic curvature is non-vanishing and the intrinsic torsion of the Carrollian structure is unconstrained. The additional fields \(b_a\) and \(S_{ab}\) become essential ingredients of the geometry. This Carrollian presentation provides the starting point for the Aristotelian and tensor-gauge descriptions discussed below.
	\end{itemize}
	
	\item \textbf{Aristotelian description.}
	For \(K\neq0\), the transformation \eqref{trnasV} shows that \(b_a\) shifts under Carroll boosts and can therefore be gauge fixed to zero. Imposing
	\begin{align}
		b_a=0
	\end{align}
	eliminates the boost redundancy and leads to an Aristotelian description in terms of the clock one-form and spatial vielbein
	\begin{align}
		(\tau_\mu,e_\mu{}^{a}),
	\end{align}
	supplemented by the symmetric tensor \(S_{ab}\).
	
	Depending on the behavior of \(\extd\tau\), the resulting geometry may be torsionless \((\tau_{\mu\nu}=0)\), twistless torsional \((\tau_{ab}=0,\tau_{0a}\neq0)\), or fully torsional \((\tau_{\mu\nu}\neq0)\).
	
	It should be emphasized that, for \(K\neq0\), this Aristotelian description is related to torsional Carroll gravity by a gauge choice. The field \(b_a\) plays the role of a St\"uckelberg field for the Carroll boost symmetry, and the two descriptions are locally gauge equivalent. This mechanism is analogous to the St\"uckelberg realization of Galilean boosts in Ho\v rava--Lifshitz gravity formulated in Newton--Cartan variables \cite{Hartong:2015zia,Afshar:2015aku}.

\item \textbf{Tensor-gauge description.}
For $K\neq0$, instead of gauge fixing the boost symmetry by setting $b_a=0$,
one may keep the boost redundancy manifest and introduce the St\"uckelberg field
$
	M_a\sim{b_a}/{K}
$.
The resulting boost-invariant combinations define an associated Aristotelian geometry while preserving the original Carroll boost symmetry.
In this description, the independent symmetric tensor field $S_{ab}$ can be reorganized together with $M_a$ and the temporal anholonomy \(\widehat{\omega}_{0a}\) into variables $(A_{ab},A_a)$ whose transformation laws are generated by the Carroll boost parameter $\lambda_a$.
This provides a tensor-gauge reinterpretation of part of the Carroll multiplet and establishes a close connection with vector-charge tensor gauge theories.
In particular, the independent field $S_{ab}$ acquires a natural gauge-theoretic interpretation. Whether the resulting structure also realizes the characteristic conservation laws and mobility restrictions of fracton systems remains an open question.

\end{enumerate}
\subsection{Tensor-gauge reinterpretation of the $K\neq0$ sector}

We now elaborate on item~3 above, which provides a tensor-gauge reinterpretation of the independent field \(S_{ab}\). As discussed previously, for \(K\neq0\) the torsional Carrollian and Aristotelian descriptions are related by a gauge choice. Rather than fixing the boost symmetry by setting \(b_a=0\), one may keep the boost redundancy manifest and introduce the St\"uckelberg field
\begin{align}
	M_a=\frac{d\,b_a}{K}\,,
\end{align}
which transforms as
\begin{align}
	\delta M_a=\lambda_a\,.
\end{align}
This field can be used to construct boost-invariant geometric combinations,
\begin{align}\label{invariant combinations}
	\widetilde\tau_\mu &= \tau_\mu - e_\mu{}^a M_a\,,
	\qquad
	\widetilde e^\mu{}_a = e^\mu{}_a + \tau^\mu M_a\,.
\end{align}
The fields \((\widetilde\tau_\mu,\widetilde e^\mu{}_a)\) are invariant under Carroll boosts and define an associated Aristotelian frame. They also preserve the orthonormality relations \eqref{orhtocomplet} and \eqref{coframe}. 

In particular, the Carroll-covariant derivative appearing in \eqref{covderivindep} can be rewritten in terms of Aristotelian quantities as
\begin{align}
	\mathcal D_a \lambda_b
	=
	\widetilde{\mathcal D}_a\lambda_b
	-
	M_a\widetilde{{\mathcal D}}_0\lambda_b
	+
	2M_{[c}K_{a|b]}\lambda_c\,,
	\label{covderlambda}
\end{align}
where \(\widetilde\partial_a=\widetilde e^\mu{}_a\partial_\mu\) and
\(
\widetilde{\mathcal D}_\mu
=
\partial_\mu-\widetilde\omega_{\mu ab}
\).
The Aristotelian spin connection
\(
\widetilde\omega_{abc}
=
\frac12
(\widetilde\Omega_{acb}
+\widetilde\Omega_{bac}
+\widetilde\Omega_{bca})
\)
is itself boost invariant.

The relation between the Carrollian and Aristotelian spin connections is
\begin{align}\label{spincontilded}
	\widehat\omega_{abc}
	=
	\widetilde{\omega}_{abc}
	-
	M_a\widehat\omega_{0bc}
	-
	2M_{[c}K_{a|b]}\,,
	\qquad
	\widehat\omega_{0ab}
	=
	\widetilde{\omega}_{0ab}\,.
\end{align}

The boost-invariant frame naturally suggests a reorganization of the independent fields of the theory.
Since the gravity multiplet contains the independent symmetric tensor \(S_{ab}\), together with the St\"uckelberg field \(M_a\), we introduce the combinations
\begin{align}
	A_a
	&=
	\widehat\omega_{0a}\,,
	\label{fractongaugfields}
	\\
	A_{ab}
	&=
	S_{ab}
	+
	\widehat\omega_{0(a}M_{b)}
	+
	\frac{1}{d}K
	\Big(
	(z-1)M_aM_b
	+
	\frac12 M\!\cdot\!M\,\delta_{ab}
	\Big)\,.
	\label{fractongaugfields2}
\end{align}

The field \(A_a\) is not an independent degree of freedom of the scaling-Carroll gravity multiplet, but is constructed from the temporal anholonomy \(\widehat\omega_{0a}\). Nevertheless, in the tensor-gauge description it plays the role of a vector gauge field, in much the same way that the Kaluza--Klein gauge field arises from an off-diagonal component of a higher-dimensional metric. Together with \(A_{ab}\), it provides a natural parametrization of the boost symmetry in terms of tensor-gauge variables.

Using \eqref{Maboost}, \eqref{indeptransf}, and the boost transformation of \(\widehat\omega_{0a}\), one finds
\begin{align}\label{fractongagugefield}
	\delta A_a
	&=
	\widetilde{\mathcal D}_0\lambda_a
	-
	K_{ab}\lambda_b\,,
	\\
	\delta A_{ab}
	&=
	\widetilde{\mathcal D}_{(a}\lambda_{b)}
	+
	K_{ab}M\!\cdot\!\lambda
	-
	2M_{(a}K_{b)c}\lambda_c\,.
\end{align}

The parameter \(\lambda_a\), originally the Carroll boost parameter, therefore acts as a vector-valued gauge parameter for the fields \(A_a\) and \(A_{ab}\).
The transformation law \eqref{fractongagugefield} has the same structural form as that encountered in vector-charge tensor gauge theories and motivates the interpretation of \(A_{ab}\) as a tensor gauge field coupled to Aristotelian geometry.

Motivated by the analogy with tensor-gauge structure, one may define the covariant combination
\begin{align}\label{fractonelectricfield}
	E_{ab}
	=
	\widetilde{\mathcal D}_0A_{ab}
	-
	\widetilde{\mathcal D}_{(a}A_{b)}
	+\cdots ,
\end{align}
where the omitted terms denote possible curvature contributions required by gauge invariance.
Similarly, in four spacetime dimensions one may introduce
\begin{align}
	B_{ab}
	=
	\varepsilon_{alm}
	\varepsilon_{bpq}
	\widetilde{\mathcal D}_m
	\widetilde{\mathcal D}_q
	A_{lp}\,,
\end{align}
which resembles the magnetic field of vector-charge tensor gauge theories \cite{Pretko:2016kxt}.

Finally, we emphasize that this construction should be viewed as a tensor-gauge reinterpretation of the \(K\neq0\) sector of scaling-Carroll gravity rather than as a separate gravitational phase. The St\"uckelberg field \(M_a\) allows one to pass between a Carrollian description and an equivalent boost-invariant Aristotelian description, while the temporal anholonomy \(\widehat{\omega}_{0a}\) and the independent tensor field \(S_{ab}\) admit a natural tensor-gauge interpretation through the variables \((A_a,A_{ab})\). The resulting gauge transformations closely resemble those of vector-charge tensor gauge theories and provide a covariant geometric framework in which such tensor-gauge structures emerge naturally from scaling-Carroll gravity for \(K\neq0\).

Establishing a direct connection with fracton physics at the dynamical level would require a further analysis of the associated currents, conservation laws, and mobility constraints, which we leave for future work.

\section{Scaling Carrollian gravity}\label{sec:Carrollian-gravity} 
In the previous section, we presented a general framework for constructing different but related gravitational descriptions starting from a scale-invariant, matter-coupled Carrollian gauge theory. In this section, we apply the conformal construction to specific scale-invariant Carrollian field theories, introducing a single scalar field $\phi$ as a compensating field for the local scaling symmetry. The subsequent gauge fixing $\phi=1$ breaks the scaling symmetry and gives rise to the Carroll-invariant gravity action. The explicit field theories and their coupling to geometry are developed in subsections \ref{sec4} and \ref{couplign}, the curvature terms are discussed in \ref{curvterms}, and the gauge fixing procedure is detailed in \ref{GFIXING}.

\subsection{Scaling-Carroll invariant field theories}
\label{sec4}
Our first goal is to classify all possible single-scalar field theories that are invariant under
the global scaling-Carroll transformations. Here we consider single real scalar field theories. The scaling dimension $w$ of the field $\phi$  transforming under a global scaling $t'=\lambda^{z}t$ and $\vec{x}'=\lambda\, \vec{x}$ is defined according to 
\begin{align}
	\phi'(t,\vec x)=\lambda^{w}\phi(\lambda^zt,\lambda\,\vec x)\,.
\end{align}

\paragraph{Carrollian supertranslation.}
One of the specific features of the Carrollian field theories is due to the presence of some supertranslaiton symmetry. A general finite Carrollian supertranslation  acts on the space and time as;
\begin{align}\label{finitesupertrans}
	\vec x'=\vec x\,,\qquad
	t'&=t-f(\vec x)\,,
\end{align}
we thus have
\begin{align}
	\begin{split}
		\partial_i'&=\partial_{i}+\partial_{i}f(\vec x)\partial_{t}\,,\qquad\partial_t'=\partial_{t}\,,\\
		\partial_i'\partial_j'&=\partial_{i}\partial_{j}+\partial_{i}\partial_{j}f(\vec x)\partial_{t}+2\partial_{(i}f(\vec x)\partial_{j)}\partial_{t}+\partial_{i}f(\vec x)\partial_{j}f(\vec x)\partial^{2}_{t} \,,\\
		\partial_i'\partial_t'&=\partial_{i}\partial_{t}+\partial_{i}f(\vec x)\partial_{t}^2  \,.
	\end{split}
\end{align}
In particular for the Carrolian boost transformation, the supertranslation function is a linear function $f(\vec x)=\beta_ix_i$ while for the temporal special conformal transformation (SCT) we have $f(\vec x)=\alpha \,x^2$. We may classify real scale invariant Carroll scalar field theories in terms of the number of their time and space derivatives. Examples of Carrolian field theories realizing the full supertransaltion symmetry is the following
\begin{subequations}\label{scalingaction0}
	\begin{align}
		\mathcal{L}_{(1,0)}&=\frac12\phi\partial_t\phi\,,\qquad w=-\frac d2 \label{scalingaction10}\\\label{scalingaction}
		\mathcal{L}_{(2,0)}&=\frac{1}{2}(\partial_{t}\phi)^{2}\,,\qquad w=\frac{z-d}{2}\,.
	\end{align}
\end{subequations}
Here $w$ is determined such that the theory is invariant under $z$-scaling. 
The one-time derivative scalar field theory with no space derivative, \eqref{scalingaction10} is trivial, since it is a boundary term. In order to make it non-trivial, we may multiply it by any boost invariant combination like $X=\partial_t\phi,\cdots$ which would effectively land us on \eqref{scalingaction} or by coupling to the gravity  curvature terms as we will see in section \ref{curvterms}. It turns out if we require to add space derivatives to the Lagrangians \eqref{scalingaction0},  apparently, the number of space derivatives should always be even in order to have rotation invariance.\footnote{Combinations like $(\partial_t\phi\partial_i\phi+\phi\partial_i\partial_t\phi)X^i$ with $X^i$ being a curvature invariant is not accepted since the expression in the parenthesis is only invariant up to total derivatives.} It is also clear that real single field, Carroll scalar theory with no time derivatives (potential term) does not exist. The one-time derivative combinations $\partial_{t}\phi\partial_{i}\phi\partial_{i}\phi+2\phi\partial_{i}\phi\partial_i\partial_t\phi$ and $\partial_{t}\phi\partial_{i}\partial_{i}\phi+\phi\partial_{i}\partial_i\partial_t\phi$ are Carroll invariant only up to a total derivative and thus they are not appropriate for coupling to gravity. 

The first non-trivial Carroll invariant combination could come with exactly two space and time derivatives;
\begin{align}
	\mathcal{L}_{(2,2)}&=\frac{1}{3}\phi[(\partial_{t}^2\phi)(\partial_{i}\partial^{i}\phi)-(\partial_{i}\partial_{t}\phi)(\partial^{i}\partial_{t}\phi)]\,,\qquad w=\frac{z+2-d}{3}\,.\label{scalingaction1}
\end{align}
The Lagrangian \eqref{scalingaction1} first appeared in \cite{Baig:2023yaz} in the context of spacetime subsystem (fractonic) symmetries. The transformation of the Lagrangian \eqref{scalingaction1} under the finite Carroll supertransaltion \eqref{finitesupertrans} is;
\begin{align}
	\mathcal{L}_{(2,2)}&\to
	\mathcal{L}_{(2,2)}+\frac13\phi\partial_t\phi\partial_t^2\phi\,\partial_i^2f(x)\,,
\end{align}
which implies that in order to have invariance we should restrict the supertranslation parameter to the case where $\partial^2f(x^i)=0$. So the scalar Lagrangian \eqref{scalingaction1}, in addition to rotation, is invariant only under a subset of supertranslations, namely Carrollian boost, 
\begin{align}
	\vec x'&=R\vec x\\
	t'&=t-\vec\beta\cdot\vec x\,.
\end{align}

Another form of classification for Carrollian field theories is referred to as the electric and magnetic versions. These two variants can be distinguished through different limiting procedures applied to the Hamiltonian of the relativistic field theories \cite{Henneaux:2021yzg}.
In the Lagrangian picture, for the electric sector one takes the limit $\epsilon\to0$ after rescaling both the field and the time coordinate $\phi\to\epsilon\,\phi$ and $t\to\epsilon \,t$ (in units where $c=1$) \cite{deBoer:2023fnj,Bergshoeff:2022qkx}. The magnetic sector limit is obtained after introducing a Lagrange multiplier $\chi$, rescaling $t\to\epsilon \,t$ and then take the strict $\epsilon\to0$ limit keeping the field $\phi$ and the Lagrange multiplier fixed. It is significant to note that in the ultimate magnetic Lagrangian, the Lagrange multiplier cannot be eliminated using its own equation of motion \cite{deBoer:2021jej,Henneaux:2021yzg}. In this sense, both field theories \eqref{scalingaction} and \eqref{scalingaction1} are classified as electric.\footnote{A relativistic origin for the Lagrangian \eqref{scalingaction1} is outlined in \cite{Kasikci:2023tvs}.}

\subsection{Scalar coupled Carroll gravities}\label{couplign}
Here, we  couple the scalar Carrollian field theories \eqref{scalingaction}-\eqref{scalingaction1} to gravity. This coupling should ensure the invariance under the whole local scaling-Carroll symmetries. 
The coupling of the above field theories to gravity is  obtained by replacing the flat space
derivatives $\partial_t$ and $\partial_i$ by covariant derivatives $D_0=\tau^\mu(\partial_\mu+\cdots)$ and $D_a=e^\mu{}_a(\partial_\mu+\cdots)$ where
the dots represent
the set of gauge fields that need to be added for covariance. Since the scalar field $\phi$ only transforms under general coordinate transformations and dilatation $\delta\phi=w\lambda_{D}\phi$,
its scaling covariant derivative is defined as follows
\begin{align}\label{1stDerv}
	D_a\phi=e^\mu{}_a\left(\partial_\mu- wb_\mu\right)\phi\,,\qquad D_0\phi=\tau^\mu\left(\partial_\mu- wb_\mu\right)\phi\,.
\end{align}
The transformation of \eqref{1stDerv} are
\begin{align}\label{transf1stD}
	\begin{split}
		{\delta}(D_{0}\phi)&=(w-z)\lambda_{D}D_{0}\phi\,\\
		{\delta}(D_{a}\phi)&=(w-1)\lambda_{D}D_{a}\phi+\lambda_{a}{}^{b}D_{b}\phi-\lambda_{a}D_{0}\phi\,.
	\end{split}
\end{align}
In order to gauge the conformal action \eqref{scalingaction} we construct the second-order derivatives from their corresponding transformation \eqref{transf1stD}
\begin{align}
	\begin{split}
		D_{0}^2\phi &=\tau^{\mu}\left(\partial_{\mu}D_{0}\phi-(w-z)b_{\mu}D_{0}\phi\right)\,,\\
		D_{a}D_{0}\phi&=e^{\mu}{}_{a}\left(\partial_{\mu}D_{0}\phi-(w-z)b_{\mu}D_{0}\phi\right)\,,\\
		D_{0}D_{a}\phi&=\tau^{\mu}\left(\partial_{\mu}D_{a}\phi-(w-1)b_{\mu}D_{a}\phi-\omega_{\mu  ab}D^{b}\phi+\omega_{\mu a}D_{0}\phi\right)\,,\\
		D_{a}D_{b}\phi&=e^{\mu}{}_{a}\left(\partial_{\mu}D_{b}\phi-(w-1)b_{\mu}D_{b}\phi-\omega_{\mu bc}D^{c}\phi+\omega_{\mu b}D_{0}\phi\right)\,.
	\end{split}
\end{align}
The gauging procedure is naturally applied by replacing ordinary derivatives with covariant derivatives in \eqref{scalingaction} and \eqref{scalingaction1};
\begin{align}\label{L02gauged}
	{\mathcal L}^{(2)}_{\text{Kin}}&=\frac{1}{2}e\,(D_0\phi)^{2}\,,\\
	{\bar{\bar{\mathcal L}}}^{(3)}_{\text{Kin}}&=\frac{1}{3}e\,\phi[(D_{0}D_{0}\phi)(D_{a}D_{a}\phi)-(D_{a}D_{0}\phi)(D_{a}D_{0}\phi)]\,.\label{L22gaugedp}
\end{align}
Interestingly, although the rigid (ungauged) field theory \eqref{scalingaction1} contains at most two time derivatives, the gauged Lagrangian \eqref{L22gaugedp} involves three time derivatives due to the presence of $D_0\phi$ inside $D_aD_a\phi$. 
In general, replacing ordinary derivatives with covariant derivatives in the field theory Lagrangian \eqref{scalingaction1} can be ambiguous for two reasons; first, because the commutation properties of partial derivatives is in general lost for the covariant derivatives. It turns out that in some
cases the covariant derivatives do not commute as will be addressed below as Ricci identities. The scaling-Carroll gauge transformations \eqref{homtransf} of the Carroll gravities \eqref{L02gauged} and \eqref{L22gaugedp} are as follows
\begin{align}
	\delta_{\text{\tiny gt}} {\mathcal L}^{(2)}_{\text{Kin}}&=(2w-z+d)\lambda_D  {\mathcal L}^{(2)}_{\text{Kin}}\,,\\
	\delta_{\text{\tiny gt}} {\bar{\bar{\mathcal L}}}^{(3)}_{\text{Kin}}&=(3w-z+d-2)\lambda_D  {\bar{\bar{\mathcal L}}}^{(3)}_{\text{Kin}}+\frac{1}{3}e\,\lambda_a\phi D_0^2\phi[D_0,D_a]\phi\,,\label{deltaL22gaugedp}
\end{align}
which fixes the scaling dimension in each cases. In order to avoid the non-invariance under boost gauge transformation in \eqref{deltaL22gaugedp} we add a supplementary term to it, which amounts to changing the order of temporal and the spatial covariant derivatives in one of the factors of the second term in \eqref{L22gaugedp}. It is precisely the following ordering that could ensure boost gauge invariance;
\begin{align}
	{\bar{{\mathcal L}}}^{(3)}_{\text{Kin}}&= {\bar{\bar{\mathcal L}}}^{(3)}_{\text{Kin}}-\frac13e\,\phi[D_0,D_a]\phi D_aD_0\phi\nn\\&=\frac{1}{3}e\,\phi[(D_{0}D_{0}\phi)(D^{a}D_{a}\phi)-(D_{a}D_{0}\phi)(D_{0}D_{a}\phi)]\label{L22gauged}
	\,.
\end{align}
Second, all forms of the Lagrangian which differ by total derivatives in the flat background cannot invariantly be coupled to gravity, not just by replacing derivatives with covariant derivatives. The reason is, after imposing the constraints and trading some gauge fields as dependent fields, the presence of the torsion in this construction could lead to non-invariance which entails adding new terms to the gravity coupled Lagrangian. We will mention this point for our case at hand as torsion identity below.

\paragraph{Ricci identities.}
Due to the presence of the torsion, in this case the covariant
derivatives do not necessarily commute on scalar fields 
\begin{align}\label{Ricciid1}
	[D_a,D_0]\phi&=-R_{a0}(H)D_0\phi-R_{a0}{}^b(P)D_b\phi-wR_{a0}(D)\phi\,,\\
	[D_a,D_b]\phi&=-R_{ab}{}^c(P)D_c\phi-R_{ab}(H)D_0\phi-wR_{ab}(D)\phi\,.\label{Ricciid2}
\end{align}
After applying the constraints which we used to solve dependent  gauge fields we have 
\begin{align}
	[D_a,D_0]\phi=\cancel{R}_{0(ab)}(P)D_b\phi-wR_{a0}(D)\phi
	\,,\qquad
	[D_a,D_b]\phi=-wR_{ab}(D)\phi\,.
\end{align}
We notice that, according to the Ricci identities \eqref{Ricciid1}, the order of the temporal and spatial covariant derivatives in the second term of \eqref{L22gauged} can lead to different results. 
\paragraph{Torsion identities.}Following our discussion in section \ref{sec.trns} we should revisit the boost invariance of the gravity coupled Lagrangian \eqref{L22gauged}. The reason is the presence of possible torsion terms which are essential to make the Lagrangian boost invariant in a curved background. 
First, we examine  the Carroll boost transformation of the covariant derivatives appearing in \eqref{L22gauged} once the spin-connections are dependent. We need to use the transformation rules  given in section \ref{sec.trns}. We have
\begin{align}
	\begin{split}
		\delta (D_{a}D_{0}\phi)&=-\lambda_{a}D^{2}_{0}\phi\,,\\
		\delta(D_{0}D_{a}\phi)&=-\lambda_{a}D^{2}_{0}\phi
		+\lambda^{b}\cancel{R}_{0(ab)}(P)D_0\phi\,,\\
		\delta(D_{a}D_{a}\phi)&=-\lambda_{a}D_{0}D_{a}\phi-\lambda_{a}D_{a}D_{0}\phi
		+\lambda_{b}\cancel{R}_{0(ab)}(P)D_a\phi\,,
		\\
		\delta(D_{a}D_{b}\phi)&=-\lambda_{a}D_{0}D_{b}\phi-\lambda_{b}D_{a}D_{0}\phi+\lambda_{b}\cancel{R}_{0(ac)}(P)D^{c}\phi-\lambda_{c}\cancel{R}_{0(ab)}(P)D^{c}\phi\,.
	\end{split}
\end{align}
Consequently, as previously mentioned, we have the following non-invariance due to the presence of the torsion:
\begin{align}\label{noninvar}
	\begin{split}
		\delta  {\bar{{\mathcal L}}}^{(3)}_{\text{Kin}}&={\frac{1}{3}}e \,\phi\left(D_{a}\phi D_{0}^{2}\phi-D_{a}D_{0}\phi {D_{0}\phi}\right)\lambda^{b}\cancel{R}_{0(ab)}(P)\,.
	\end{split}
\end{align}
In order to guarantee the invariance of the Lagrangian under Carroll boost transformations, it is essential to introduce the following supplementary terms; 
\begin{align}\label{XXXX}
	{{{\mathcal L}}}^{(3)}_{\text{Kin}}= {\bar{{\mathcal L}}}^{(3)}_{\text{Kin}}+{\frac{1}{3}}e\,\phi\Big(D_{0}D_{a}\phi D_{b}\phi-D_{a}D_{b}\phi D_{0}\phi+\frac{1}{2} D_{c}\phi D_{c}\phi
	\cancel{R}{}_{0(ab)}(P)\Big)\cancel{R}{}_{0(ab)}(P)\,.
\end{align}
One can show that the transformation of the supplementary term cancels out the non-invariance appearing in \eqref{noninvar}, such that   the improved Lagrangian $  {{{\mathcal L}}}^{(3)}_{\text{Kin}}= {\bar{{\mathcal L}}}^{(3)}_{\text{Kin}}+eX$ is invariant $\delta  {{{\mathcal L}}}^{(3)}_{\text{Kin}}=0$.

\subsection{Curvature terms}\label{curvterms}
There exist Carroll-invariant curvature-term Lagrangians that have no field-theoretic counterpart. In other words, they do not arise from coupling a scalar field theory to gravity. Nevertheless, such curvature terms can be added with arbitrary coefficients to the kinetic terms discussed above. An important point to address is the number of derivatives that these invariants are allowed to contain. To determine this, we focus on the number of time derivatives appearing in the kinetic Lagrangian \eqref{XXXX}. As a consequence,  the most general Carroll-invariant curvature contributions to the Lagrangian \eqref{XXXX} may include terms with both two and three time derivatives.

Those curvature invariants containing second-order time derivatives take the general form
\begin{equation}
	\label{curv2time}
	e^{-1}\mathcal{L}^{(2)}_{\text{curv}} = \alpha_1\, R(G,J) +\alpha_2\, \cancel{R}^2_{0}{}_{(ab)}(P) + \alpha_3\, R_{0a}^2(D) \,,
\end{equation}
where $R^2_{ab}$ and $R_{0a}^2$ refer to curvature squared given in \eqref{curvatureinv1} while $R(G,J)$ is given in \eqref{curvatureinv2}. They are  expressed in terms of independent variables;
\begin{align}
	\cancel{R}{}^{2}_{0}{}_{(ab)}(P)&=K_{ab}K^{ab}-\frac1dK^2\,,\\
	R_{0a}^2(D)&=(\mathcal{D}_{0}b_{a}-{K_{ab}b_b}
	+\frac1d\partial_{a}K)^2\,,\\
	R(G,J)&=R(G,J)\Big\rvert_{S_{ab}=0}-K_{ab}S^{ab}+\partial_{0}S +\frac{z-1}{d}SK\,,
\end{align}
where we replaced  the dependent gauge field $b_0$ and other dependent fields from \eqref{b0gaugefield} and \eqref{torsion} in terms of the extrinsic curvature. Furthermore, 
\begin{align}
	R(G,J)\Big\rvert_{S_{ab}=0}=2R_{0a}{}^a(G)\Big\rvert_{S_{ab}=0}+R_{ab}{}^{ab}(J)
\end{align}
where 
\begin{align}
	R_{0a}{}^a(G)\Big\rvert_{S_{ab}=0}&=-\partial_a\omega_{0a}+\Omega_{aba}\omega_{0b}+\omega_{0a}\omega_{0a}+(d-2)b_b\omega_{0b}\nn\\
	&=-{\mathcal D}_a\tau_{0a}-z{\mathcal D}\cdot b+\tau_{0a}\tau_{0a}+(2z+d-2)\tau_{0a}b_a+z(z+d-2)b\cdot b\,,
\end{align}
where the Carrollian covariant derivatives in the second line are defined as in \eqref{covderivindep} for $S_{ab}=0$.
It is noticeable that the independent field $S_{ab}$ appear only in the curvature terms through $R(G,J)$. 

The curvature terms including third-order time derivatives are given by
\begin{equation}
	\label{curv3time}
	e^{-1}\mathcal{L}^{(3)}_{\text{curv}} = \beta_1\,\cancel{R}^3_{0(ab)}(P) + \beta_2\, {R}_{0a}(D){R}_{0b}(D)\cancel{R}_{0}{}^{(ab)}(P)+ \cdots\, ,
\end{equation}
where $R^3_{ab}$ refers to the curvature cubed $R_{ac}R_{cd}R_{da}$ and dots denotes all possible independent scalar combinations built from $\mathcal L^{(2)}_{\text{curv}}$ and $\phi\,D_0\phi$ as the gauged Lagrangian \eqref{scalingaction10}
constructing up to three time derivatives. 
We may continue this construction with up to 4 time derivatives with
\begin{align}
	\label{curv4time}
	e^{-1}\mathcal{L}^{(4)}_{\text{curv}} =\gamma_1\,\cancel{R}^4_{0(ab)}(P)+\cdots\,,
\end{align}
where the dots denote additional possible invariants with four time derivatives constructed from lower derivative invariants in the
gauged Lagrangian \eqref{L02gauged} and \eqref{curv2time}. 

In summary, these curvature contributions must be included in the most general Carroll-invariant action for a scalar field coupled to background geometry, with their coefficients left arbitrary at this stage.

\subsection{Gauge fixing}\label{GFIXING}
We are ready to implement the appropriate dilatation gauge fixing to obtain Carroll invariants.
Setting \(\phi = 1\) in the Lagrangian density \eqref{L02gauged} we have
\begin{align}\label{L20gaugefixed}
	\mathcal{L}^{(2)}_{\text{Kin}}&=\frac{w}{2d}\,e\,\Big( \partial_{0}K+\frac{w-z}{d}K^{2}\Big)\,.
\end{align}
Note that each term in the above is independently Carroll invariant independent of the value of the weight $w$. In fact as discussed below eq. \eqref{Kinv}, if we perform partial integration both terms lead to a same invariant $K^2$.

Accordingly, the gravitational theory derived from the combination of ${\mathcal{L}}^{(3)}_{\text{Kin}}$ in 
\eqref{XXXX} is expressed as follows:
\begin{align}\label{L22gaugedfixed}
	\begin{split}
		{\mathcal{L}}^{(3)}_{\text{Kin}}=&\frac{e}{3d}w^{2}\bigg\{\Big[-(\partial_{0}K)(\mathcal{D}\cdot b)+(\partial^{a}K)(\mathcal{D}_{0}b_{a}) \Big]-(w-z)\Big[\frac{1}{d}K^{2}(\mathcal{D}\cdot b)+Kb^{a}(\mathcal{D}_{0}b_{a}) \Big]\\
		&+(w-1)\Big[\frac{1}{d}K\ b_{a}(\partial^{a}K)+b\cdot b\ \partial_{0}K\Big]\\
		&+(d-1)b\cdot b\,\partial_{0}K-\frac zd K\,b_{a}(\partial^{a}K)+(w-z)(d+z-1)\frac1d K^{2}\,b\cdot b \\&+d\Big(b_{a}\mathcal{D}_{0}b_{b}-(z-1)\frac{1}{d}K\,b_{a}b_{b}-\frac{1}{d}K\,b\cdot b\,\delta_{ab}+\frac{1}{d}K\mathcal{D}_{a}b_{b} \Big)\Big(-K^{ab}+\frac{1}{d}\delta^{ab}K \Big)\\&+\frac{d}{2}b\cdot b\Big(K_{ab}K^{ab}-\frac1dK^2\Big)\bigg\}\,.
	\end{split}
\end{align}
The definition of the covariant terms $\mathcal D_0b_a$ and $\mathcal D_ab_b$ (covariant with respect to Carroll boosts
and spatial rotations) is given in \eqref{covderivindep}. The Carroll invariance of this gravitational Lagrangian is directly checked in appendix \ref{AppDinvar}. As in \eqref{L20gaugefixed}, the invariance holds for all values of $w$ for a fixed $z$, and thus we are left with two independent invariants: 
\begin{align}\label{L22gaugedfixed5}
	I_1&=-(\partial_{0}K)(\mathcal{D}\cdot b)+(\partial^{a}K)(\mathcal{D}_{0}b_{a}) -\frac{1}{d}K\ b_{a}(\partial^{a}K)+(d+z-2)b\cdot b\ \partial_{0}K\nn\\&+d\Big(b_{a}\mathcal{D}_{0}b_{b}-(z-1)\frac{1}{d}Kb_{a}b_{b}-\frac{1}{d}K\,b\cdot b\,\delta_{ab}+\frac{1}{d}K\mathcal{D}_{a}b_{b} \Big)\Big(-K^{ab}+\frac{1}{d}\delta^{ab}K \Big)\nn\\&+\frac{d}{2}b\cdot b\Big(K_{ab}K^{ab}-\frac1dK^2\Big)\,,\\
	I_2&= -\frac{1}{d}K^{2}(\mathcal{D}\cdot b)-Kb^{a}(\mathcal{D}_{0}b_{a}) +\frac{1}{d}K\ b_{a}(\partial^{a}K)+b\cdot b\ \partial_{0}K+(d+z-1)\frac1d K^{2}\,b\cdot b \,.\label{L22gaugedfixed6}
\end{align}
The invariant Lagrangians \eqref{L22gaugedfixed5} and \eqref{L22gaugedfixed6} have up to three time derivatives. This can be easily checked from the coefficient of the independent field $S_{ab}$ which appears as a Lagrange multiplier in $\mathcal D_ab_b$ terms.  
If we vary the general Lagrangian $I_1+\zeta \,I_2$ w.r.t. $S_{ab}$ we get the following constraint:
\begin{align}\label{extrinsiccurvequ}
	(K\partial_0 K+\frac{\zeta-1}{d}K^3)\delta_{ab}+K^2K_{ab}=0\,,
\end{align}
where $\zeta=w-z$ is an arbitrary parameter.
This equation leads to the trivial solution  $K_{ab}=0$  and a non-trivial solution which, assuming $\tau^\mu=(1,\vec 0)$ can be expressed as
\begin{align}\label{extrinsicsolu}
	K_{ab}=\frac{1}{C(\vec x)+\zeta\,t}\delta_{ab}\,.
\end{align}
where $C(\vec x)$ is an arbitrary scalar function of spatial coordinates. The $t$-dependence of the extrinsic curvature, shows its dynamical evolution over time.  

It would be interesting to check the invariants in two limiting cases $z\to0,\infty$ where the scaling from the field theory point of view is purely spatial ($t\to t$ and $\vec x\to\lambda\vec x$) or temporal ($t\to\lambda t$ and $\vec x\to\vec x$), respectively. The solution \eqref{extrinsicsolu} in these cases leads to 
\begin{align}
	z\to0\,:\quad K_{ab}\to \frac{1}{C(\vec x)+w \,t}\delta_{ab}\,,\quad\text{and}\quad z\to\infty\,:\quad K_{ab}\to-\frac1zt^{-1}\delta_{ab}\to0\,,
\end{align}
where $w$ is a free parameter.

At this level we can apply the procedure of section \ref{sec:conformal-construction} to the invariant $I_1 +\zeta\,I_2$. Corresponding to this invariant we have following cases.
\paragraph{\textbf{Sheared Carrollian Gravity.}}
By enforcing the constraint \(K=0\), the resulting model reduces to a sheared Carrollian gravity theory in the terminology of section \ref{sec:conformal-construction}, and can be represented as follows:
\begin{align}
	-K^{ab} \ b_{a}\mathcal{D}_{0}b_{b}+\frac12 b\cdot b\,K_{ab}K^{ab} \,.
\end{align}
Here the Carrollian covariant derivative is defined as in \eqref{covderivindep} for $K=0$.

\paragraph{\textbf{Aristotelian Gravity.}}
In this case the vector field \( b_{a} \) vanishes, while \( K \neq 0 \). As a result, the Carrollian boost symmetry is broken and the associated gravitational theory is Aristotelian. We have:
\begin{align}
	SK(\partial_{0}K)-\widehat\omega_{0a}K(\partial_{a}K)-\frac{1}{d}S_{ab}K^{2}\left (-K^{ab}+\frac{1}{d}\delta^{ab}K\right) +\frac{\zeta}{d}SK^{3}\,.
\end{align} 
The temporal torsion in this geometric setting is represented by \(\widehat{\omega}_{0a} = \tau_{0a}\). 
The Aristotelian transformations act on the tensor \( S_{ab} \) as:
\begin{align}
	\delta S_{ab}=2\lambda_{(a}{}^{c}\ S_{b)c}+\xi^\mu\partial_\mu S_{ab}\,.
\end{align}

\paragraph{\textbf{Tensor Gauge Theory.}}\label{fractongravity}
In this case following the procedure in section \ref{sec:conformal-construction}, we first compensate for boost transformation by  redefining the gauge field \(\tau_{\mu}\) and the covector \(e^{\mu}{}_{a}\) and covariant derivatives as in \eqref{invariant combinations}-\eqref{spincontilded}.
Upon implementing these redefinitions, the invariants $I_1$ and $I_2$ turn into the following form:
\begin{align}\label{fractonLagrang}
	\begin{split}
		I_1\quad\to\quad&\frac{1}{d}K\partial_0K\left(A_{aa}-\widetilde{\mathcal{D}}\cdot M-\frac 12M\cdot MK+K^{ab}M_{a}M_{b}\right)\\
		+&\frac{1}{d}K\widetilde{\partial}_aK\left(-A_{a}+\widetilde{\mathcal{D}}_0 M_a-K_{ab}M_{b}\right)\\
		+&\frac{1}{d}K^2\left(-A_{ab}+\widetilde{\mathcal{D}}_a M_b+\frac 12M\cdot MK_{ab}-K_{ac}M_{b}M_{c}\right)\left(-K_{ab}+\frac1dK\delta_{ab}\right)
		\,,\\
		I_2\quad\to\quad&\frac{1}{d^2}K^3\left(A_{aa}-\widetilde{\mathcal{D}}\cdot M-\frac 12M\cdot MK+K^{ab}M_{a}M_{b}\right)\,.
	\end{split}
\end{align}
In these Lagrangians \eqref{fractonLagrang}, all covariant derivatives appearing above are Aristotelian;
\begin{align}
	\widetilde{\mathcal{D}}_a M_b=\widetilde{\partial}_aM_b-\widetilde{\omega}_{abc}M_c\,,\qquad\quad
	\widetilde{\mathcal{D}}_0 M_a={\partial}_0M_a-\widehat{\omega}_{0ab}M_b\,.
\end{align} One can observe the emergence of gauge fields \( A_{ab} \) and \( A_{a} \) as fracton gauge fields and \( M_{a} \) as a vector and the gauge invariance of fracton  Lagrangians \eqref{fractonLagrang} under gauge transformation given in \eqref{Maboost} and \eqref{fractongagugefield}. We can do the following field redefinition in the introduced gauge fields \eqref{fractongaugfields2};
\begin{align}
	A_a&\;\to\; A_a-K_{ab}M_b\,,\\
	A_{ab}&\;\to\;A_{ab}+\frac12M\cdot MK_{ab}-K_{(a|c}M_{b)}M_c\,,
\end{align}
such that the gauge transformation \eqref{fractongagugefield} change  to
\begin{align}
	\delta A_a=\widetilde{\mathcal D}_0\lambda_a\,,\quad\qquad%
	\delta A_{ab} = \widetilde{\mathcal D}_{(a}\lambda_{b)}-M_{(a}K_{b)c}\lambda_c+M_cK_{(a|c}\lambda_{b)}\,,
\end{align}
and the Lagrangians \eqref{fractonLagrang} simplify to 
\begin{align}
	\begin{split}
		I_1\quad\to\quad&\frac{1}{d}K\partial_0K\left(A_{aa}-\widetilde{\mathcal{D}}\cdot M\right)
		+\frac{1}{d}K\widetilde{\partial}_aK\left(-A_{a}+\widetilde{\mathcal{D}}_0 M_a\right)\\
		&+\frac{1}{d}K^2\left(-A_{ab}+\widetilde{\mathcal{D}}_a M_b\right)\left(-K_{ab}+\frac1dK\delta_{ab}\right)
		\,,\\
		I_2\quad\to\quad&\frac{1}{d^2}K^3\left(A_{aa}-\widetilde{\mathcal{D}}\cdot M\right)\,.
	\end{split}
\end{align}
Thus we are left with three gauge invariant scalar combination of  gauge fields
\begin{align}
	A_{aa}-\widetilde{\mathcal{D}}\cdot M\,,\qquad \left(A_{ab}-\widetilde{\mathcal{D}}_a M_b\right)K_{ab}\,,\qquad \widetilde{\partial}_aK\left(A_{a}-\widetilde{\mathcal{D}}_0 M_a\right)\,.
\end{align}

\section{Conclusion}\label{sec:Conclusion}

In this work, we formulated matter-coupled scaling-Carroll gravity as a gauge theory based on a compensating real scalar field with a general dynamical exponent $z$. Our construction extends the existing construction of Carroll gravity at $z=1$ in which invariance under Carrollian special conformal symmetry is also demanded \cite{Bergshoeff:2024ilz}. Within our framework, Carrollian scalar field theories remain invariant under local Carroll and anisotropic scaling transformation. 

One outcome of our analysis is that the extrinsic curvature $K_{ab}$ is no longer forced to vanish by the equations of motion of the Lagrange multiplier $S_{ab}$ (originating from the Carroll boost connection) \cite{Bergshoeff:2017btm,Campoleoni:2022ebj}. This becomes possible because special conformal symmetry is relaxed and the dilatation gauge sector contributes an additional spatial vector field $b_a$ to the multiplet. After fixing the scaling symmetry, this vector acquires a shift transformation under local Carroll boosts proportional to the trace of the extrinsic curvature $K$. The fields $(\tau_\mu,e_\mu{}^{a},b_a,S_{ab})$ then admit different related geometric descriptions depending on how the boost symmetry is treated.

When the Carroll boost symmetry is left unfixed, the theory describes
\emph{Carroll gravity}, characterized by a genuinely Carrollian geometry
with non-vanishing extrinsic curvature. This description admits both a
subsector with vanishing trace $K=0$, governed by a Carrollian shear tensor, and a torsional sector with $K\neq0$, in which the intrinsic torsion is unconstrained.

It should be emphasized that the Carrollian sectors obtained in our 
construction are not identical to the standard Carroll gravity theories
discussed in the literature. In both cases, the geometry is supplemented by an additional symmetric tensor field $S_{ab}$ originating from the Carroll boost connection.

For $K\neq0$, fixing the boost symmetry by imposing $b_a=0$ leads to an
\emph{Aristotelian description} of the same underlying theory. In this case, the independent fields reduce to the Aristotelian clock one-form and spatial vielbein, together with the symmetric tensor $S_{ab}$. Depending on the behavior of the temporal Aristotelian torsion $\extd\tau$, the resulting geometry can be torsionless, twistless torsional, or torsional in the language of non-relativistic geometry \cite{Christensen:2013lma}.

Finally, when both $b_a$ and the trace of the extrinsic curvature are
non-vanishing and the boost symmetry remains unfixed, the boost-shifting
vector $b_a$ can be used to construct boost-invariant geometric data. The
resulting boost-invariant Aristotelian description admits a tensor-gauge
reinterpretation in terms of fields $(A_a,A_{ab})$, where the independent
symmetric tensor $S_{ab}$ naturally acquires the role of a tensor gauge
field, while the temporal anholonomy $\widehat{\omega}_{0a}$ provides the
associated vector gauge field needed to construct gauge-invariant
combinations. In this description, the Carroll boost parameter plays the
role of a vector-valued gauge parameter. The resulting gauge
transformations closely resemble those of vector-charge tensor gauge
theories on an Aristotelian background.

These descriptions illustrate the rich interplay between Carrollian, Aristotelian, and tensor-gauge structures encoded in a common set of geometric variables. Our construction provides a unified framework in which they are not separate theories but arise as different realizations of the same underlying scaling-Carroll gauge structure, as summarized in Fig.~\ref{fig1} and Table~\ref{tab:comparison}. An important open question is whether the tensor-gauge structure identified here gives rise to the characteristic conservation laws and mobility constraints associated with fracton systems. Addressing this issue requires coupling the theory to suitable matter sources and analyzing the corresponding currents and Ward identities. We hope to return to these questions in future work.

\begin{table}[t]
	\centering
	\renewcommand{\arraystretch}{1.2}
	\begin{tabular}{|l|c|c|c|}
		\hline
		Description & Boosts & $b_a$ & Fields \\
		\hline\hline
	
		Sheared Carroll gravity ($K=0$)
		& Unfixed
		& Physical
		& $(\tau_\mu,e_\mu{}^{a},b_a,S_{ab})$
		\\ \hline
		
		Aristotelian description ($K\neq0$)
		& Gauge-fixed
		& $0$
		& $(\tau_\mu,e_\mu{}^{a},S_{ab})$
		\\ \hline
		
		Tensor-gauge description ($K\neq0$)
		& Reinterpreted
		& $M_a\sim b_a/K$
		& $(\widetilde\tau_\mu,\widetilde e_\mu{}^{a},A_a,A_{ab})$
		\\ \hline
		
	\end{tabular}
\caption{
	Geometric descriptions arising in scaling--Carroll gravity.
	The $K=0$ sector corresponds to sheared Carroll gravity.
	For $K\neq0$ sector, corresponding to the torsional Carroll, one may either gauge-fix the boost and obtain an Aristotelian description, or retain the boost redundancy and reinterpret part of the gravity multiplet in terms of tensor-gauge variables.
}
	\label{tab:comparison}	
\end{table}

Several directions for future work follow from our results. These include applying a same conformal program to the Galilei case \cite{Afshar:2015aku,Abedini:2019voz,Devecioglu:2018apj}, supersymmetric extensions of scaling-Carroll gravity via the gauging of Carroll superalgebras \cite{Bergshoeff:2015wma,Bagchi:2022owq}, as well as applications to flat-space holography \cite{Donnay:2023mrd,Adami:2023wbe,Bagchi:2023cen}, Carrollian hydrodynamics \cite{Ciambelli:2018wre}, effective geometric descriptions of fractonic matter coupled to  curved spacetime \cite{Jain:2021ibh,Hartong:2024hvs,Afxonidis:2025wce,Pena-Benitez:2023aat,Fecit:2025eet} and Carrollian analogues of Horndeski-type theories \cite{Kasikci:2023tvs}.

\acknowledgments
We thank Mojtaba Najafizadeh and Reza Yousofi for discussions. We also thank the anonymous referee for his/her valuable comments. MAJ additionally thanks the participants of the IPM HEP-TH weekly group meetings for valuable discussions on related topics. This work is based upon research funded by Iran National Science Foundation (INSF) under project No. 4044075.
\appendix

\section{Identities}\label{appA}

The  Carrollian  frame fields $\tau_{\mu}$, $e_{\mu}{}^{a}$ and their inverses have the following transformaiton under Carrolian boost\begin{align}\label{boosttrans}
	\delta_{{\text{\tiny G}}} e_\mu{}^a=0=\delta_{{\text{\tiny G}}}\tau^{\mu}\,,\qquad \delta_{{\text{\tiny G}}} e^\mu{}_a=-\lambda_a\tau^\mu\,,  \qquad  \delta_{{\text{\tiny G}}} \tau_\mu&=\lambda_{a}e_{\mu}{}^{a}\,.
\end{align}
These frame fields define a non-singular $D\times D$ matrix $(\tau_{\mu},e_{\mu}{}^{a})$ 
with following non-vanishin determinant,  
\begin{align}
	e\equiv \det(\tau_{\mu},e_{\mu}{}^{a})=\sqrt{\det\left(\tau_{\mu}\tau_{\nu}+h_{\mu\nu} \right)}\,,
\end{align}
which is invariant under Carroll boost transformations;
\begin{align}
	\delta_{G}\, e= \frac{1}{2} e\left(\tau^{\mu}\tau^{\nu}+h^{\mu\nu} \right)\delta_{G} \left(\tau_{\mu}\tau_{\nu}+h_{\mu\nu}\right)
	=0\,.
\end{align}
The partial derivative of the determinant is 
\begin{align}
	\partial_{\mu}e
	=e\,(\tau^{\nu}\partial_{\mu}\tau_{\nu}+e^{\nu}{}_{a}\partial_{\mu}e_{\nu}{}^{a})\,.
\end{align}
The derivative of the vector $\tau^\mu$ and the inverse Vielbein can be calculated in terms of form fields,
\begin{align}
	\partial_{\mu}\tau^{\rho}&=-\tau^{\nu}\tau^{\rho}\partial_{\mu}\tau_{\nu}-\tau^\nu  e^{\rho}{}_{a} \partial_{\mu} e_{\nu}{}^a \,,\\ 
	\partial_{\mu}e^{\nu}{}_{a}&=-\tau^{\nu}e^{\rho}{}_{a}\partial_{\mu}\tau_{\rho}-e^{\rho}{}_{a}e^{\nu}{}_{b}\partial_{\mu}e_{\rho}{}^{b}\,.
\end{align}
Using the above identities, we have the following:
\begin{align}
	e^{-1}\partial_{\mu}(e\,\tau^{\mu})&=e^{-1}\tau^{\mu}\partial_{\mu}e+\partial_{\mu}\tau^{\mu}
	\\&=2\tau^{\mu}\ e^{\nu}{}_{a} \partial_{[\mu}\ e_{\nu]}{}^{ a}=-K\,,\\
	e^{-1}\partial_{\mu}(e\, e^{\mu}_{\ a})&=2e^{\mu}_{\ a}\ e^{\nu}_{\ b}\partial_{[\mu}e_{\nu]}^{\ b}-e^{\nu}_{\ a}\tau^{\mu}\partial_{[\mu}\tau_{\nu]}\nn\\
	&=\Omega_{ab}{}^{b}-\tau_{0a}\,,
\end{align}
where we also used the definition of the extrinsic curvature and anholonomy coefficients introduced in section \ref{sec:Carrollian-geometry}.
\section{Transformation rules}\label{trnsrules}
It is essential to obtain the transformation rules of dependent fields and compare with their transformation as independent gauge fields mentioned in \eqref{homtransf}. We can show that these two transformations are not necessarily the same and can be different by some amount corresponding to the unconstrained torsion. For example, the gauge transformation of the spin-connection as an independent gauge field comes directly from \eqref{homtransf}:
\begin{align}\label{trans0omega}
	\begin{split}
		\delta_{\text{\tiny{gt}}} \omega_{[ab]}
		&=-\lambda_{[a}\tau_{0b]}+\partial_{[a}\lambda_{b]}-\frac12\lambda_{cb}\tau_{ac}+\frac12\lambda_{ca}\tau_{bc}+\frac12\lambda_c\Omega_{abc}+\frac12(z-2)\lambda_D\tau_{ab}\,,\\
		\delta_{\text{\tiny{gt}}}\omega_{0a}&=\partial_{0}\lambda_{a}-\lambda_D\tau_{0a}-z\lambda_Db_a+\lambda_{ab}\tau_{0b}+z\lambda_{ab}b_b+\lambda_{b}\Omega_{0}{}_{[ab]}-(1-z)\frac1d\lambda_{a}K\,,\\
		\delta_{\text{\tiny{gt}}}\omega_{abc}&=-\lambda_D\omega_{abc}+\lambda_{ad}\omega_{dbc}+\lambda_{bd}\omega_{adc}-\lambda_{cd}\omega_{adb}+\partial_a\lambda_{bc}+\lambda_{a}\Omega_{0 [bc]}\,,\\
		\delta_{\text{\tiny{gt}}}\omega_{0ab}&=\partial_0\lambda_{ab}-\lambda_{ac}\Omega_{0[cb]}+\lambda_{bc}\Omega_{0[ca]}+z\lambda_D\Omega_{0[ab]}\,.
	\end{split}
\end{align}
We can also obtain the  transformation of these fields as dependent fields appearing in \eqref{boostspin} and \eqref{rotationspin};
\begin{align}
	\begin{split}
		\delta \omega_{[ab]}(e,\tau)
		&=-\lambda_{[a}\tau_{0b]}+\partial_{[a}\lambda_{b]}+\frac{1}{2}\lambda_{c}\Omega_{ab}{}^{c}+\frac12\lambda_{bc}\tau_{ac}+\frac12\lambda_{ac}\tau_{cb}+\frac12(z-2)\lambda_D\tau_{ab} \,,\\
		\delta \omega_{0a}(e,\tau)
		&=\partial_{0}\lambda_{a}-\lambda_D\tau_{0a}-z\lambda_Db_a+\lambda_{ab}\tau_{0b}+z\lambda_{ab}b_b+\lambda_{b}\Omega_{0}{}_{ab}+\frac{z}{d}\lambda_{a}K\,,\\
		\delta\omega_{abc}(e,\tau,b_a)&=\lambda_{a}\Omega_{0[bc]}-\lambda_{b}({\Omega_{0(ca)}}-b_{0}\delta_{ca})+\lambda_{c}({\Omega_{0(ab)}}-b_{0}\delta_{ba})+\cdots \,,\\
		\delta\omega_{0ab}(e,\tau,b_a)&=\partial_0\lambda_{ab}-\lambda_{ac}\Omega_{0[cb]}+\lambda_{bc}\Omega_{0[ca]}+z\lambda_D\Omega_{0[ab]}\,.
	\end{split}
\end{align}
where we used the fact that
$\delta \Omega_{abc}=\cdots-\lambda_a\Omega_{0bc}-\lambda_b\Omega_{a0c}$. The dots are referring to transformation under rotation and scaling which is the same as the associated gauge transformaiton.
By comparing we realize that 
\begin{align}\label{deltadleta}
	\begin{split}
		\Delta \omega_{[ab]}&=\delta \omega_{[ab]}-\delta_{\text{\tiny{gt}}} \omega_{[ab]}=0\,,\\
		\Delta \omega_{0\,a}&=\delta\omega_{0\,a}-\delta_{\text{\tiny{gt}}}\omega_{0\, a}=-\lambda_{a}b_{0}{+\lambda_b\Omega_{0(ab)}}\,,\\ \Delta\omega_{abc}&=\delta\omega_{abc}-\delta_{\text{\tiny{gt}}}\omega_{abc}=-\lambda_{b}({\Omega_{0(ca)}}-b_{0}\delta_{ca})+\lambda_{c}({\Omega_{0(ab)}}-b_{0}\delta_{ba})\, ,\\
		\Delta\omega_{0\,ab}&=\delta\omega_{0\,ab}-\delta_{\text{\tiny{gt}}}\omega_{0\,ab}=0\,.
	\end{split}
\end{align}
The Carrollian boost transformation of the dilatation gauge field \( b_{\mu} \) is zero, both as a dependent field and as an independent gauge field, since $\delta b_0=0$.

The Carrollian spin-connection $\widehat\omega$ are naturally defined in terms of the scaling-Carroll spin-connection $\omega$, addressed in section \ref{section3}, once the dilatation gauge field $b_\mu$ is zero;
\begin{align}
	\widehat\omega_{\mu}{}^a=\omega_{\mu}{}^a\bigg\rvert_{b_{\mu}=0}\,,\qquad\qquad    \widehat\omega_{\mu}{}^{ab}=\omega_{\mu}{}^{ab}\bigg\rvert_{b_{\mu}=0}\,.
\end{align}
Using the transformation rules \eqref{homtransf} and \eqref{trans0omega}, It is easy to find the boost gauge transformation of the Carrollian spin-connection in our setup:
\begin{align}\label{covtrans0}
	\delta_{\text{\tiny{gt}}}\widehat\omega_{0a}&=\partial_0\lambda_a-\widehat\omega_{0ab}\lambda_b-\frac1d\lambda_aK\,,\\
	\delta_{\text{\tiny{gt}}}\widehat\omega_{abc}&=-\lambda_a\widehat\omega_{0bc}-\frac1d K(\lambda_c \delta_{ab}-\lambda_b\delta_{ac})\,,\\
	\delta_{\text{\tiny{gt}}}\widehat\omega_{ab}&=\partial_a\lambda_b-\widehat\omega_{abc}\lambda_c-\widehat\omega_{0b}\lambda_a+2(1-z)\lambda_{(a}b_{b)}-\lambda\cdot b\delta_{ab}\,.\label{covtrans20}
\end{align}
These transformation \eqref{covtrans0}-\eqref{covtrans20} coincide with the usual Carroll boost transformation in the Carroll gauging algebra once $b_\mu=0$. 
Now, since the gauge fields \(\widehat\omega_{\mu}{}^{a}\) and \(\widehat\omega_{\mu}{}^{ab}\) are dependent,  we have 
\begin{align}
	\begin{split}
		{\Delta\widehat\omega_{0a}}&=-\lambda_{ b}K_{ab} +{\frac{1}{d}}\lambda_{ a}K, \\
		{\Delta\widehat\omega_{aa}{}_b}&=\lambda_{a}K_{ab}-{\frac1d}\lambda_{b}K,\\
		{\Delta\widehat\omega_{ab}{}_c}&=2\lambda_{[b}K_{ac]}{-\frac1d(\lambda_b \delta_{ac}-\lambda_c\delta_{ab})K}.
	\end{split}
\end{align}

The deviation of the transformation of the spin-connection under boost in \eqref{deltadleta} from standard gauge transformation, naturally propagates into curvature 2-forms.
We may analyze the transformation behavior of curvature 2-forms associated to scaling-Carroll gravity  under Carrollian boost transformations.
In particular, when the gauge fields \(\omega_{\mu}{}^{a}\) and \(\omega_{\mu}{}^{ab}\) are regarded as independent, the Carrollian boost transformation of the curvature 2-forms are 
\begin{align}\label{gtindepend}
	\begin{split}
		\delta_{\text{\tiny{gt}}} R_{\mu\nu}{}^{a}(P)&=0\,,\\
		\delta_{\text{\tiny{gt}}} R_{\mu\nu}{}^{a}(G)&=-\lambda_{b}R_{\mu\nu}{}^{ab}(J)+(1-z)\lambda^aR_{\mu\nu}(D)\,,\\
		\delta_{\text{\tiny{gt}}} R_{\mu\nu}{}^{ab}(J)&=0\,,\\
		\delta_{\text{\tiny{gt}}} R_{\mu\nu}(D)&=0\,.
	\end{split}
\end{align}
When gauge fields $\omega_{\mu}{}^{a}$ and $\omega_{\mu}{}^{ab}$ are dependent fields, the transformation of curvature 2-forms $ R_{\mu\nu}{}^{a}(P)$ and $R_{\mu\nu}(D)$ remain the same as \eqref{gtindepend}. We thus have
\begin{align}\label{gtindepend1}
	\begin{split}
		\delta R_{0a}(D)&=\delta_{\text{\tiny{gt}}} R_{0a}(D)=0\,,\\
		\delta R_{ab}(D)&=\delta_{\text{\tiny{gt}}} R_{ab}(D)=-2\lambda_{[a}R_{0|b]}(D)\,,\\
		\delta R_{0\, ab}(P)&=    \delta_{\text{\tiny{gt}}} R_{0\,ab}(P)=0\,,
	\end{split}
\end{align}
where we used the fact that $R_{0\,ab}(P)=2\tau^{\mu}e^{\nu}_{a}\partial_{[\mu}e_{\nu]\ b}-\omega_{0\, ba}-b_{0}\delta_{ba}$. 
The boost transformation of the curvature 2-form $R_{\mu\nu}{}^{a}(G)$ when gauge fields $\omega_{\mu}{}^{a}$ and $\omega_{\mu}{}^{ab}$ are dependent, is 
\begin{align}
	\delta R_{\mu\nu}{}^{a}(G)=\delta_{\text{\tiny{gt}}}R_{\mu\nu}{}^{a}(G)+\Delta R_{\mu\nu}{}^{a}(G)
\end{align}
where $\Delta R_{\mu\nu}{}^{a}(G)=2\partial_{[\mu}\Delta\omega_{\nu]}{}^{a}-2\Delta\omega_{[\mu}{}^{ab} \omega_{\nu]b}-2\omega_{[\mu}{}^{ab}\Delta\omega_{\nu]b}+2(z-1)\Delta\omega_{[\mu}{}^{a}b_{\nu]}$ and using \eqref{deltadleta} we have $\Delta \omega_{\mu}{}_a=\tau_{\mu}\Delta\omega_{0}{}_a$ and $\Delta\omega_{\mu}{}^{ab}=e_{\mu}{}_c\Delta\omega^{cab}$. In particular,  we have:
\begin{align}\label{deltagg}
	\Delta R_{0a}{}^{b}(G)=R_{0a}(H)\lambda_{c}\cancel{R}_{0}{}^{(bc)}(P)-D_{a}(\lambda_{c}\cancel{R}_{0}{}^{(bc)}(P))+b_{a}\lambda_{c}\cancel{R}_{0}{}^{(bc)}(P)\,.
\end{align}
In deriving eq. \eqref{deltagg} we used the expression for $R_{0a}(H)=2\tau^{\mu}e^{\nu}{}_{a} \partial_{[\mu}\tau_{\nu]}+(z-1)b_{a}-\omega_{0}{}_a$. 
The covariant derivative on the right hand side of \eqref{deltagg} is defined as
\begin{align}
	{D_{a}(\lambda_{b}\cancel{R}_{0}{}^{(ab)}(P))}&= D_{a}\lambda_{b}\cancel{R}_{0}{}^{(ab)}(P)+\lambda_{b}D_{a}(\cancel{R}_{0}{}^{(ab)}(P))\nn\\
	&={\partial_{a}(\lambda_{b}\cancel{R}_{0}{}^{(ab)}(P))-\omega_{a}{}^{a}{}_{c}\lambda_{b}\cancel{R}_{0}{}^{(bc)}(P) +b_a(\lambda_{b}\cancel{R}_{0}{}^{(ab)}(P))}\,,
\end{align}
where we used the fact that $D_{a}\lambda_{b}=\partial_a\lambda_b-\lambda_c\omega_{abc}-(z-1)b_a\lambda_b$ and that
\begin{align}
	D_\mu\left(\cancel{R}_{0}{}^{(ab)}(P)\right)=\partial_\mu\left(\cancel{R}_{0}{}^{(ab)}(P)\right)-\omega_\mu{}^{ac}\cancel{R}_{0}{}^{(cb)}(P)-\omega_\mu{}^{bc}\cancel{R}_{0}{}^{(ac)}(P)+z\,b_{\mu}\cancel{R}_{0}{}^{(ab)}(P)\,.
\end{align}
This is defined due to the transformation $\delta_{\text{\tiny{gt}}}\cancel{R}_{0}{}^{(ab)}(P)=\lambda_{ac}\cancel{R}_{0}{}^{(cb)}(P)+\lambda_{bc}\cancel{R}_{0}{}^{(ac)}(P)-z\lambda_D\cancel{R}_{0}{}^{(ab)}(P)$ and the fact that $\Delta \cancel{R}_{0}{}^{(ab)}(P)=0$.

The  boost transformation of the rotation curvature 2-form when gauge fields $\omega_{\mu}{}^{a}$ and $\omega_{\mu}{}^{ab}$ are dependent is
\begin{align}
	\delta R_{\mu\nu}{}^{ab}(J)=\delta_{\text{\tiny{gt}}}R_{\mu\nu}{}^{ab}(J)+\Delta R_{\mu\nu}{}^{ab}(J)\,,
\end{align}
where  $\Delta R_{\mu\nu}{}^{ab}(J)=2\partial_{[\mu}\Delta\omega_{\nu]}{}^{ab}-2\Delta\omega_{[\mu}{}^{c[a}\omega_{\nu]}{}^{b]}{}_c-2\omega_{[\mu}{}^{c[a}\Delta\omega_{\nu]}{}^{b]}{}_c$.
As a consequence we have 
\begin{align}
	{\Delta} R_{ {ab}}{}^{ab}(J)&=R_{ab}{}^c(P)\Delta\omega^c{}_{ab}+\omega_a{}^c{}_b\Delta\omega^c{}_{ab}-\omega_b{}^c{}_a\Delta\omega^c{}_{ab}-b_b\Delta\omega^a{}_{ab}+b_a\Delta\omega^b{}_{ab}\nn\\&+\partial_{a}\Delta\omega^{bab}-\partial_{b}\Delta\omega^{aab}-\Delta\omega_a{}^{ca}\omega_b{}^{bc}-\omega_a{}^{ca}\Delta\omega_b{}^{bc}+\Delta\omega_a{}^{cb}\omega_b{}^{ac}+\omega_a{}^{cb}\Delta\omega_b{}^{ac}\nn\\
	&=   {R_{abc}(P)\lambda_{b}\cancel{R}_{0}{}^{(ac)}(P)+2D_a\left(\lambda_b\cancel{R}_{0(ab)}(P)\right)-2b_a\lambda_b\cancel{R}_{0(ab)}(P)}
	\,,
\end{align}
where we used the fact that $    R_{ab}{}^c(P)=2e^{\mu}{}_{a}e^{\nu}{}_{b}\ \partial_{[\mu}e_{\nu]}{}^{c}-\omega_a{}^c{}_b+\omega_b{}^c{}_a+\delta_{ac}b_b-\delta_{bc}b_a$.

\section{Gravitational Carroll invariance}\label{AppDinvar}
Using \eqref{covtrans0} we have the following boost gauge transformation on covariant derivatives:
\begin{align}
	\delta_{\text{\tiny{gt}}}(\partial_{0}K)&=0\,,\\
	\delta_{\text{\tiny{gt}}}(\partial_{a}K)&=
	-\lambda_{a}\partial_{0}K ,\\
	\delta_{\text{\tiny{gt}}}(\mathcal{D}_{0}b_{a})&=
	\frac1d\lambda_{a}\partial_{0}K{+\frac{1}{d^2}\lambda_{a}K^{2}}. \\
	\delta_{\text{\tiny{gt}}}(\mathcal{D}\cdot b)&=
	-\lambda_{a}\mathcal{D}_{0}b_{a}+\frac1d \lambda_{a}\partial_{a}K{-\frac1d(3-2z-2d)\lambda\cdot bK}, \\
	\delta_{\text{\tiny{gt}}}(\mathcal{D}_{a}b_{b})&=
	-\lambda_{a}\mathcal{D}_{0}b_{b}+\frac1d \lambda_{b}\partial_{a}K{-\frac1d\left(-2\lambda\cdot b\delta_{ab}+2(1-z)\lambda_{(a}b_{b)}+b_a\lambda_b\right)K}\,.
\end{align}
We can check the boost gauge transformation ($\delta_{\text{\tiny{gt}}}$) of the first three lines in \eqref{L22gaugedfixed}:
\begin{align}
	\delta_{\text{\tiny{gt}}} \bar{\mathcal L}^{(3)}_{\text{Kin}}&=\frac{1}{d^2}\lambda_{a}\partial^{a}K\,K^{2}+\frac1d(3-2z-2d)\partial_0K\lambda\cdot bK\nn\\&-(w-z)\left(\frac{1}{d^2}K^2\lambda_{a}\partial_{a}K +\frac1dKb_a\lambda_{a}\partial_{0}K-\frac{2}{d^2}(1-z-d)K^3\lambda\cdot b\right)\nn\\&+(w-1)\left(\frac{1}{d^2}K^2\lambda_a (\partial^{a}K)+\frac1db^{a}\lambda_{a}K\ \partial_{0}K\right)\\&+\frac2d(d-1)\lambda\cdot bK\partial_0K-\frac{z}{d^2}K^2\lambda_a\partial_aK+\frac{z}{d}b\cdot\lambda K\partial_0K+(w-z)(d+z-1)\frac{2}{d^2}K^3\lambda\cdot b \,.\nn
\end{align}
which is zero as we expected.
This, however, will not hold once we trade the spin-connections as dependent fields which is our case. 
Now, since the gauge fields \(\widehat\omega_{\mu}{}^{a}\) and \(\widehat\omega_{\mu}{}^{ab}\) are dependent,  we have 
\begin{align}
	\Delta(\partial_{0}K)&=0,\\
	\Delta(\partial_{a}K)&=0,\\        \Delta(\mathcal{D}_{0}b_{a})&={\Delta\widehat\omega_{0a}b_0}={\frac1d}\lambda_{ b}K_{ab} K-{\frac{1}{d^2}}\lambda_{ a}K^{ 2}, \\
	\Delta(\mathcal{D}_{a}b_{a})&={-\Delta\widehat\omega_{aa}{}^bb_b}=-\lambda_{a}K_{ab}b^{b}+{\frac1d}\lambda_{b}b^{b}K,\\
	\Delta(\mathcal{D}_{a}b_{b})&={-\Delta\widehat\omega_{ab}{}^cb_c}=-2\lambda_{[b}K_{ac]}b^{c}{+\frac1d(K\lambda_b b_a-\lambda\cdot b\delta_{ab}K)}.
\end{align}
In particular, the $\Delta$ transformation of the first term in the squared bracket of eq. \eqref{L22gaugedfixed} is non-zero and
thus implementing the $\Delta$ transformation on the first three lines in \eqref{L22gaugedfixed} we get
\begin{align}
	\Delta \bar{\mathcal L}^{(3)}_{\text{Kin}}=-\partial_0K\left(-\lambda_{a}K_{ab}b^{b}+{\frac1d}\lambda_{b}b^{b}K\right)+\partial_aK\left(\frac1d\lambda_{ b}K_{ab} K-{\frac{1}{d^2}}\lambda_{ a}K^{ 2}\right)
\end{align}
On the other hand the total boost transformation ($\delta=\delta_{\text{\tiny{gt}}}+\Delta$) of the last two lines in  \eqref{L22gaugedfixed} gives
\begin{align}\label{varL22}
	&\;\;\;\;d\delta\Big(b_{a}\mathcal{D}_{0}b_{b}-(z-1)\frac{1}{d}Kb_{a}b_{b}-\frac{1}{d}Kb\cdot b\delta_{ab}+\frac{1}{d}K\mathcal{D}_{a}b_{b} \Big)\left(-K^{ab}+\frac{1}{d}\delta^{ab}K \right)\nn\\&
	\;\;\;+{d}\delta b_{c}b^{c}\big(K_{ab}K^{ab}-\frac1dK^2\big)\nn\\&=
	d\bigg(-\lambda_{a}b_0\mathcal{D}_{0}b_{b}+b_{a}(\frac1d\lambda_b\partial_0K+\frac1d\lambda^cKK_{bc})+(z-1)\frac{1}{d}K(\lambda_{a}b_{b}+b_{a}\lambda_{b})b_0\nn\\&+\frac{2}{d}K\lambda\cdot bb_0\delta_{ab}+\frac{1}{d}K\Big[-\lambda_{a}\mathcal{D}_{0}b_{b}+\frac{1}{d}\lambda_{b}\partial_{a}K-\lambda_{b}K_{ac}b^{c}+\lambda\cdot b(K_{ab}+\frac1d\delta_{ab}K)\nn\\&-\frac2d(1-z)\lambda_{(a}b_{b)}K\Big]\bigg)
	\left(-K^{ab}+\frac{1}{d}\delta^{ab}K \right)-{d}\lambda\cdot b b_0\bigg(K_{ab}K^{ab}-\frac1dK^2\bigg)\\&=
	\bigg(b_{a}(\lambda_b\partial_0K+\lambda^cKK_{bc})+K\Big[\frac{1}{d}\lambda_{b}\partial_a K-\lambda_{b}K_{ac}b^{c}\Big]\bigg)
	\left(-K^{ab}+\frac{1}{d}\delta^{ab}K \right)=-\Delta  \bar{\mathcal L}^{(3)}_{\text{Kin}}\,.\nn
\end{align}
Which shows that the total Lagrangian is invariant under Carrolian boost.
In deriving \eqref{varL22} we used the  total form of the transformation under Carrollian boost 
\begin{align}
	\begin{split}
		\delta(\partial_{0}K)&=0,\\
		\delta(\partial_{a}K)&=-\lambda_{a}\partial_{0}K,\\
		\delta(\mathcal{D}_{0}b_{a})
		&=\frac{1}{d}\lambda_{a}\partial_{0}K+\frac{1}{d}\lambda^{b}KK_{ab}, \\
		\delta(\mathcal{D}\cdot b)
		&=\frac{1}{d}\lambda_{a}\partial_{a}K-\lambda_{a}\mathcal{D}_{0}b_{a}-\lambda_{a}K_{ab}b^{b}+{\frac{2}{d}(z+d-1)\lambda\cdot bK},  \\
		\delta(\mathcal{D}_{a}b_{b})
		&=-\lambda_{a}\mathcal{D}_{0}b_{b}+\frac{1}{d}\lambda_{b}\partial_{a}K-\lambda_{b}K_{ac}b^{c}+\lambda\cdot b\big(K_{ab}+\frac1d\delta_{ab}K\big)-\frac2d(1-z)\lambda_{(a}b_{b)}K
	\end{split}
\end{align}


\addcontentsline{toc}{section}{References}
\bibliographystyle{fullsort.bst}
\bibliography{biblio}

\end{document}